\documentclass[12pt,notitlepage,a4paper]{article}
\pdfoutput=1
\usepackage[a4paper,margin=1in]{geometry}
\usepackage{amsmath,amssymb,amscd,amsfonts}
\usepackage{mathtools}
\numberwithin{equation}{section}
\usepackage{url}
\usepackage[shortlabels]{enumitem}
\usepackage{tikz} 
\usetikzlibrary{decorations}
\usetikzlibrary{arrows,decorations.markings}         
\usepackage[hidelinks]{hyperref}
\usepackage[open,
  openlevel=2,
  atend,
  numbered]{bookmark}
\usepackage{graphicx}
\usepackage[final]{pdfpages}
\usepackage{threeparttable}
\usepackage{caption}
\usepackage{float}
\usepackage{stackengine}
\usepackage{scalerel}
\usepackage[bf]{titlesec}
\titlespacing{\section}{3pc}{2pc}{0.8pc}
\titleformat{\section}
  {\centering\large\bfseries} 
  {\Large\thesection}
  {0.5em} 
  {\Large}  
\titlespacing{\subsection}{0pc}{2pc}{0.8pc}
\titleformat{\subsection}
  {\centering\bfseries} 
  {\normalsize\thesubsection}
  {0.5em} 
  {\normalsize}  
\usepackage{tocloft}

\newlength{\mylen}	
\urldef\footurlb\url{cocalc.com/dfriedan/DM/SM}
\urldef\footurla\url{physics.rutgers.edu/~friedan}
\def\eq{\begin{equation}}
\def\en{\end{equation}}
\def\eqg{\eq\begin{gathered}}
\def\eng{\end{gathered}\en}
\def\eqa{\eq\begin{aligned}}
\def\ena{\end{aligned}\en}
\def\Reals{\mathbb{R}}

\def\SO{\mathrm{SO}}
\def\SU{\mathrm{SU}}

\def\Spin{\mathrm{Spin}}
\DeclareMathOperator{\cn}{cn}

\def\expval#1{\langle \, #1 \,\rangle}
\def\Vol{\mathrm{Vol}}
\def\gauge{\mathrm{gauge}}
\def\HO{{\scriptscriptstyle\mathrm{HO}}}

\def\tr{\mathrm{tr}}
\def\Complexes{\mathbb{C}}
\def\Integers{\mathbb{Z}}

\def\cQ{\mathcal{Q}}
\def\cA{\mathcal{A}}
\def\cW{\mathcal{W}}
\def\cP{\mathcal{P}}
\def\cO{\mathcal{O}}
\def\cG{\mathcal{G}}

\def\cV{\mathcal{V}}
\def\cM{\mathcal{M}}
\def\cR{\mathcal{R}}
\def\cD{\mathcal{D}}
\def\cH{\mathcal{H}}
\def\cJ{\mathcal{J}}
\def\cB{\mathcal{B}}
\def\cF{\mathcal{F}}
\def\bK{\mathbf{K}}

\def\bbT{\mathbb{T}}
\def\bbL{\mathbb{L}}
\def\Moni{\cM_{\mathrm{i}}}

\def\kB{k_{\scriptscriptstyle \mathrm{B}}}

\def\phys{\scriptstyle \mathrm{phys}}

\def\CGF{\scriptscriptstyle \mathrm{CGF}}

\def\epsilonb{\epsilon}

\def\ECGF{E_{\CGF}}

\def\gtwo{g}

\stackMath
\def\dyhat{-0.2ex}
\newcommand\myhat[1]{\ThisStyle{%
              \stackon[\dyhat]{\SavedStyle#1}
                              {\SavedStyle\hat{\phantom{#1}}}}}
\def\that{\kern0.1em\myhat{\kern-0.1em t}}

\def\texttilde{\raise-0.7ex\hbox{\!\texttt{\char`\~}}}
\def\starnabla{{*}\hat\nabla}
\def\asymp{\mathrm{asymp}}
\usepackage{afterpage}
\begin{document}
\def\title{Thermodynamic stability of a cosmological\\[1ex]
SU(2)-weak gauge field}
\begin{center}
{\LARGE \title}
\vskip4ex
{\large Daniel Friedan}
\vskip2ex
{\it
New High Energy Theory Center
and Department of Physics and Astronomy,\\
Rutgers, The State University of New Jersey,\\
Piscataway, New Jersey 08854-8019 U.S.A. and
\vskip1ex
Science Institute, The University of Iceland,
Reykjavik, Iceland
\vskip1ex
\href{mailto:dfriedan@gmail.com}{dfriedan@gmail.com}
\qquad
\href{https://physics.rutgers.edu/\textasciitilde friedan/}
{physics.rutgers.edu/\texttilde friedan}
}
\vskip2ex
March 22, 2022
\end{center}
%
%
%
\begin{center}
\vskip3ex
{\sc Abstract}
\vskip2.5ex
\parbox{0.96\linewidth}{
\hspace*{1.5em}
The CGF cosmology is a complete theory 
of cosmology
from the electroweak transition onward.
It is semi-classical.
At leading order the only matter is dark matter
--- a cosmological SU(2)-weak gauge field (the CGF).
Ordinary matter is a subleading correction 
from fluctuations around the classical state.
The CGF is periodic in imaginary time.
It acts as thermal bath for the fluctuations
of the Standard Model fields.
Here, the initial thermal state of the 
$\SU(2)$ gauge field fluctuations is constructed
and shown to be thermodynamically stable.
This is a warm-up for (1) constructing the initial thermal state of all the 
fluctuations in order to calculate its time evolution
and (2) showing that initial state to be thermodynamically stable
in order to show that the CGF cosmology is physically natural.
}
\end{center}

\vspace*{-2ex}

%
%
%
\begin{center}
\tableofcontents
\end{center}
%
%
\section{Introduction}

The CGF cosmology is a complete theory
of the Standard Model cosmological epoch,
from the electroweak transition onward
\cite{Friedan:2020poe,Friedan2022:Atheory,Friedan2022:DMStars}.
The theory has no free parameters 
and  assumes no physical laws beyond the Standard Model and General Relativity.
All of cosmology is given by the time evolution of
a uniquely determined highly symmetric semi-classical initial state
in the period leading up to the electroweak transition.
The CGF universe in the leading order,
classical approximation contains only a cosmological SU(2)-weak gauge field (the CGF).
The CGF is the dark matter.
The relatively small amount of ordinary matter in the universe is a 
higher order correction to the dark matter universe
from the fluctuations of the Standard Model 
fields around the classical CGF.

The CGF cosmology is completely determined by four assumptions.
\begin{enumerate}
\item The universe is governed by the Standard Model
and General Relativity (with cosmological constant).
Nothing beyond the Standard Model,
nothing beyond the known laws of physics, is assumed.
\item The universe is a 3-sphere.
\item The state of the universe is 
invariant under a $\Spin(4)$ symmetry group
that acts on the 3-sphere as $\SO(4)$
and on the Standard Model fields
such that the $\SU(2)$-weak doublets transform as spinors.
\item The initial energy in the Standard Model fields is $ >  10^{107}$ in natural units.
\end{enumerate}
The $\Spin(4)$ symmetry and the initial energy completely
determine the classical initial condition.
The only nontrivial $\Spin(4)$-symmetric Standard Model field is the SU(2)-weak gauge 
field, the CGF.
It is a classical solution of the Yang-Mills equation of motion
given by an elliptic function of the complex time.
It is periodic in real time, oscillating anharmonically in the quartic Yang-Mills action.
It is also periodic in imaginary time.
The periodicity in imaginary time defines a temperature.
The initial state of the fluctuations is
given by correlation functions that respect 
the imaginary time periodicity.
The CGF acts 
as a thermal bath for the fluctuations 
of the Standard Model fields.

Here, the initial thermal state of the SU(2)-weak gauge field
fluctuations is constructed.
This is done by by constructing the 2-point function
in the gaussian approximation.
The $n$-point functions are Wick contractions
of the 2-point function.
Thenthe thermal state is
shown to be thermodynamically stable by a combination of mathematical proof and
numerical evidence.

The space of fluctuations of the $\SU(2)$ gauge field around the classical
solution is decomposed under the $\Spin(4)$ symmetry group.  The
fluctuations in each irreducible representation satisfy, in the
gaussian approximation, an
equation of motion that is a
linear ordinary differential equation (ode) second order in the time variable.
The coefficients of the ode
are analytic in the time variable.
A certain property {\bf P} of the ode is proved to imply the existence of a 
canonical change of variables mapping the thermal state of the gauge field 
fluctuations to an equilibrium state of an ordinary harmonic 
oscillator.  So property {\bf P} implies thermodynamic stability.
Strong numerical evidence is given that property {\bf P} holds
and thus that the initial thermal state of the gauge field 
fluctuations is thermodynamically stable.
Thermodynamic stability means that
the cosmological initial condition is robust against 
small fluctuations of the $\SU(2)$ gauge field.

Calculations are shown in the Supplemental Material 
\cite{Friedan2022:StabilitySuppMat}.
The numerical calculations are done in SageMath \cite{sagemath9.4}
using the mpmath arbitrary-precision floating-point arithmetic 
library \cite{mpmath}.


\section{Spin(4)-symmetric CGF}

Space is a 3-sphere.
Let $S^{3}$ be the unit 3-sphere in $\Reals^{4}$
with O(4)-symmetric metric
$\hat g_{ij}(\hat x)$ 
and volume 3-form $\hat \epsilon_{ijk}(\hat x)$.
Let $\gamma_{i}(\hat x)$ be the Spin(4)-symmetric Dirac matrices on $S^{3}$
and let $\hat \nabla_{i}$ be the Spin(4)-symmetric
covariant derivative on spinors (and tensors).
\eqg
\hat \gamma_{i}\hat \gamma_{j} = -\frac14 \hat g_{ij} -\frac12 
\hat \epsilon_{ij}{}^{k}\gamma_{k}
\qquad
\hat \nabla_{i}\hat \gamma_{j}  =0
\qquad
[\hat \nabla_{i},\,\hat  \nabla_{j}] = \epsilon_{ij}{}^{k} \hat \gamma_{k}
\eng
The O(4)-symmetric space-time metric (in $c=1$ units) is
\eq
ds^{2} = R(\that)^{2}\left(
-d\that{\;\!}^{2} +  \hat g_{ij}(\hat x)d\hat x^{i}d\hat x^{j}
\right)
\en
$\that$ is conformal cosmological time.
$R(\that)$ is the radius of the spatial 
3-sphere at conformal time $\that$.
Co-moving time $t$ is given by
$dt = R(\that) d \that$.

The SU(2)-weak gauge bundle over the 3-sphere is identified with the
spinor bundle.
The general Spin(4)-symmetric gauge field in unitary gauge is
\eqg
D_{0} = \partial_{\that}
\qquad
D_{i} = \hat \nabla_{i} + \hat b(\that) \hat \gamma_{i}
\\[1ex]
F_{0j} = [D_{0},\,D_{j}]  = \frac{d\hat b}{d\that} \hat \gamma_{j}
\qquad
F_{ij} = [D_{i},\,D_{j}] =  (1-b^{2})\hat \epsilon_{ij}{}^{k}\hat \gamma_{k}
\eng
The Yang-Mills action is
\eq
\label{eq:YMaction}
\frac1\hbar S_{\gauge} =\int
\frac1{2 \gtwo ^{2}}
\tr(- F_{\mu\nu} F^{\mu\nu})
\, \sqrt{-g} \,d^{4}x
\en
$g$ is the SU(2) coupling constant of the Standard Model.
The Yang-Mills action is conformally invariant so is
independent of the space-time scale $\hat R$.
The action of the $\Spin(4)$-symmetric gauge field is
\eq
\label{eq:bhataction}
\frac1\hbar S_{\gauge} =
\Vol(S^{3})\frac{3  }{g^{2}}
\int
\bigg[-\frac12 \bigg(\frac{d\hat b}{d\that}\bigg)^{2}+ \frac12 
(\hat b^{2}-1)^{2}\bigg]
d\that
\qquad
\Vol(S^{3}) = 2\pi^{2}
\en
The equation of motion is
\eq
\frac{d^{2}\hat b}{d\that^{2}} + 2 \hat b (\hat b^{2}-1) = 0
\en
The dimensionless energy
\eq
\ECGF = \frac12 \left(\frac{d\hat b}{d\that}\right)^{2} + \frac12(\hat b^{2}-1)^{2}
\en
is conserved.
The solution (up to time translation) is the elliptic function
\eqg
\label{eq:classicalsoln}
\hat b(\that) = \frac{1}{\epsilonb } k\cn( z,k)
\qquad
\ECGF = \frac1{8\epsilonb ^{4}}
\qquad
z = \frac{1}{\epsilonb } \that
\qquad
k^{2}= \frac12 + \epsilonb ^{2}
\eng
This is the initial cosmological gauge field,
prior to the electroweak transition.
The observed flatness of the present universe
requires 
$\ECGF > 10^{107}$, $\epsilonb < 10^{-27}$
\cite{Friedan2022:Atheory}.

\section{The Jacobi elliptic function $\cn(z,k)$}

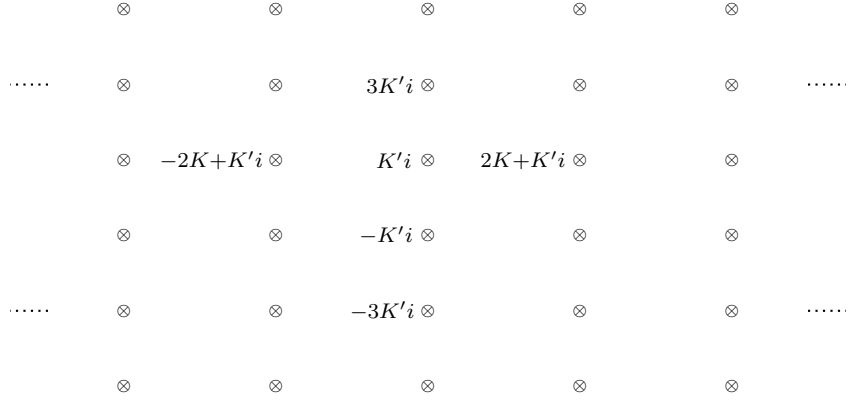
\begin{figure}
\begin{center}
\begin{tikzpicture}[baseline=2ex,x=0.5cm,y=0.5cm]
%
\foreach \x in {-4,0,4,8,12}
{
\node at (\x,5) {$\scriptscriptstyle \otimes$};
\node at (\x,3) {$\scriptscriptstyle \otimes$};
\node at (\x,1) {$\scriptscriptstyle \otimes$};
\node at (\x,-1) {$\scriptscriptstyle \otimes$};
\node at (\x,-3) {$\scriptscriptstyle \otimes$};
\node at (\x,-5) {$\scriptscriptstyle \otimes$};
}
\node at (4-0.9,1){$\scriptstyle K'i$};
\node at (4-1,3){$\scriptstyle 3K'i$};
\node at (4-1.1,-1){$\scriptstyle -K'i$};
\node at (4-1.2,-3){$\scriptstyle -3K'i$};
\node at (8-1.5,1){$\scriptstyle 2K+K'i$};
\node at (-1.7,1){$\scriptstyle -2K+K'i$};
\draw[dotted,thick] (14,3) -- (15,3);
\draw[dotted,thick] (14,-3) -- (15,-3);
\draw[dotted,thick] (-6,3) -- (-7,3);
\draw[dotted,thick] (-6,-3) -- (-7,-3);
\end{tikzpicture}
\end{center}
\caption{The poles of $\cn(z,k)$ are at $z_{m,n}=2 m K + (2n+1) K' i$.  
}
\label{fig:poles}
\end{figure}

$\cn(z,k)$ is a Jacobi elliptic function \cite[Chapter 22]{DLMF2022},
\cite[8.14-15]{GR7th}.
It is analytic in $z$ (with poles).
It satisfies
\eqg
\label{eq:cnderivatives}
\cn'' = ( k^{2}-k'{}^{2})  \cn-2k^{2} \cn^{3}
\qquad
\cn'{}^{2} = (1-\cn^{2})(k'{}^{2}+k^{2}\cn^{2})
\\[1ex]
k^{2}+k'{}^{2} =1
\eng
$k^{2}$ is called the {\it parameter}, 
$k'{}^{2}$ the {\it complementary parameter},
$k$ and $k'$ the {\it modulus} and {\it complementary modulus}.
We assume  $0<k,\,k'<1$.
We are especially interested in $k^{2} = \frac12 +\epsilon^{2}$,
$k'{}^{2} = \frac12 -\epsilon^{2}$ with 
$\epsilon \ll 1$.
The Taylor series at $0$ is
\eq
\cn(z) = 1 -\frac12 z^{2} + O(z^{4})
\en
The reflection symmetries are
\eqg
\cn(\bar z) = \overline{\cn(z)}
\qquad
\cn(-z) = \cn(z)
\eng
$\cn(z,k)$ is doubly periodic in $z$.
\eqg
\cn(z) = \cn(z+4K) = \cn(z+4K'i) = \cn(z+2K +2K'i)
\\[1ex]
K =K(k)
\qquad
K' = K(k')
\eng
$K$ and $K'$ are the complete elliptic integrals of the first kind.
The half-periods are
\eq
\cn(z+2K,k) = \cn(z+2K'i,k) = - \cn(z,k)
\en
The poles and residues are
\eq
\cn(z,k) \sim    \frac{(-1)^{m+n+1} i k^{-1}}{z-z_{m,n}}
\qquad
z_{m,n} = 2 m K + (2n+1) K' i
\qquad
m,\,n\in \Integers
\en
The poles are shown in Figure~\ref{fig:poles}.
The zeros are located by the identity
\eq
\cn(z+K+K'i) = \frac{-ik'k^{-1}}{\cn(z)}
\en

\section{Periodicity in imaginary time as temperature}

The CGF $\hat b (\that)$ oscillates anharmonically with period 
$\Delta \that = 4K\epsilon$.  The period in co-moving time
is $\Delta t = 4K \epsilonb  R(\that)$.
The CGF is also periodic in imaginary time with period $\Delta t = 
4K' \epsilonb R(\that) i$, defining a temperature $T_{\CGF}(\that)$ 
by
\eq
\frac{\hbar}{\kB T_{\CGF}(\that)} = 4K' \epsilonb R(\that)
\en
The initial state of the fluctuations is determined by the 
periodicity in imaginary time.
In the path integral formulation of the Standard Model,
the CGF is a classical trajectory in the phase space of SU(2) gauge 
theory.  The classical trajectory analytically continues to an 
analytic trajectory in the complexified phase space.
The initial thermal state of the fluctuations is defined by the path 
integral over the paths periodic in imaginary time.

This is a familiar picture when the classical real time trajectory is 
invariant under time translation.  Here the classical CGF is not invariant 
under time translation.  Moreover, the analytic continuation in 
complex time is 
obstructed by poles of the classical trajectory.
It must be proved that the thermal state of the fluctuations 
is independent of the choice of periodic path in imaginary time over which the functional 
integral is performed.

\section{Stability}

After the CGF cosmological initial condition was proposed in \cite{Friedan:2020poe},
the stability of the CGF was investigated in \cite{Kumar:2021vdm}.
Stability is a crucial physical requirement.
An instability would render
implausible that the initial condition could
result from earlier cosmological developments.
A stable initial condition 
is robust against small fluctuations ---
the essential condition of physical naturalness. 
However,
the stability investigated in \cite{Kumar:2021vdm} 
was stability
of the classical gauge field
under small classical perturbations.
Classical stability is not the relevant physical stability condition
for the 
cosmological initial condition.
The initial condition is 
a \emph{semi}-classical thermal {quantum} state,
i.e.~a state concentrated near a classical trajectory
that is periodic in imaginary time.
The physical stability condition is {thermodynamic stability}.
The thermal state has to be constructed and
then shown to be thermodynamically stable.

Thermodynamic stability is the condition that 
the gaussian approximation to the path integral 
should be well defined
---
the quadratic approximation to the imaginary time action should be 
bounded below.
Equivalently, 
the gaussian approximation to
the quantum mechanical density matrix
should be positive definite.
Here, thermodynamic stability is shown by constructing a canonical 
transformation to an 
equilibrium thermal state of an ordinary harmonic oscillator.

\section{Quadratic term in the action}

The gauge field fluctuations are the perturbations $B_{i}(\hat x)$ of the 
classical solution,
\eq
\tilde D_{i} =  \hat \nabla_{i} + \hat b \gamma_{i} + 
B_{i}
\qquad
B_{i}(\hat x) = B_{i}^{j}(\hat x)\gamma_{j}(\hat x)
\en
modulo the infinitesimal gauge transformations
\eq
B_{i}^{\gauge} =  \hat \nabla_{i}v + \hat b [\gamma_{i},\,v]
\qquad
v(\hat x) = v^{j}(\hat x)\gamma_{j}(\hat x)
\en
The gaussian path integral is constructed
from the quadratic term in the action (\ref{eq:YMaction}).
Change time variable from conformal time $\that$ to 
$z$.
\eq
z =\frac{\that}{\epsilon}
\qquad
\hat b(\that) = \frac{b(z)}{\epsilon}
\qquad
b(z) = k \cn(z,k)
\en
and define two linear operators on the perturbations,
\eqg
\label{eq:Gammastarnabla}
\Gamma B_{i} = \epsilon_{i}{}^{jk}[\gamma_{j},\, B_{k}]
\qquad
\starnabla B_{i} = \epsilon_{i}{}^{jk}\hat \nabla_{j} B_{k}
\eng
After some algebra, the quadratic term in the action is
written \cite{Friedan2022:StabilitySuppMat}
\eqg
\label{eq:quadraticYMaction}
\frac{1}\hbar S_{2} 
=\frac1 {\epsilon g ^{2}} \int -\tr \left(
- \partial_{z}B^{i}\partial_{z} B_{i}
+B^{i} \bK(z) B_{i}\right)
\, \sqrt{-\hat g} \,d^{3}\hat x
\; dz
\\[2ex]
\bK(z) 
=  b(z)^{2}\bK_{2}  +  b(z)\bK_{1} 
+ \bK_{0}  
\\[2ex]
\bK_{2} =  \Gamma^{2}-\Gamma
\qquad
\frac1{\epsilon} \bK_{1} = \Gamma (\starnabla)+(\starnabla)\Gamma
\qquad
\frac1{\epsilon^{2}} \bK_{0} =  (\starnabla )^{2}+\Gamma 
\eng
The operators $\Gamma$ and $\starnabla$ are Spin(4)-invariant.
When the space of perturbations $B_{i}^{j}(\hat x)$ is decomposed
under Spin(4), 
the operator $\bK(z)$ becomes block diagonal.
The gaussian quantum field theory becomes a discrete sum of 
finite quantum mechanical systems.

\section{Spin(4) decomposition}

Identify $S^{3}$ with the group $\SU(2)$.
Then $\Spin(4)$ is $\SU(2)_{L}{\times}\SU(2)_{R}$ acting by left 
and right multiplication on $\SU(2)$.
Write the irreducible representations of SU(2) in the usual way
$j =  0,\; 1/2,\; 1,\; 3/2,\; \ldots$ with $\dim(j) = 2j+1$.
The irreducible representations of Spin(4) are the tensor products
$(j_{L},j_{R})$ of an $\SU(2)_{L}$ irreducible with an $\SU(2)_{R}$
irreducible.
The space of functions on 
$S^{3}$ decomposes under $\Spin(4)$ as
the representation $\mathop\oplus_{j} (j,j)$.
Identify the tangent and cotangent spaces of $S^{3}$  with the Lie algebra of 
$SU(2)_{L}$ which is the representation $(1,0)$.
The space of gauge field fluctuations decomposes as
\eq
\label{eq:Bdecomp}
\left\{ B_{i}^{j}(\hat x)\right\} 
=  \mathop\oplus\limits_{j_{R}} \;(1\otimes 1\otimes j_{R}, j_{R})
= \mathop\oplus\limits_{j_{L},j_{R}} \; 
\Complexes^{N(j_{L},j_{R})}\otimes (j_{L},j_{R})
\en
The multiplicity $N(j_{L},j_{R})$ is 0, 1, 2, or 3.
The space of infinitesimal gauge transformations decomposes as
\eq
\left\{ v^{j}(\hat x)\right\} =  \mathop\oplus\limits_{j_{R}} 
\;(1 \otimes j_{R}, j_{R})
= \mathop\oplus\limits_{j_{L},j_{R}} \; 
\Complexes^{N_{\gauge}(j_{L},j_{R})}\otimes (j_{L},j_{R})
\en
The multiplicity $N_{\gauge}(j_{L},j_{R})$ is 0 or 1.
The multiplicity of physical degrees of freedom is $N_{\mathrm{phys}} = N -  
N_{\mathrm{gauge}}$.
The representations with $N_{\mathrm{phys}}>0$ fall into 
the five subsets listed in Table~\ref{table-irreps}.
\begin{table}
$$
\begin{array}{c@{\qquad}c@{\qquad}c@{\qquad}cc@{\qquad}cc@{\qquad}c}
 & j_{L}-j_{R} &(j_{L},j_{R}) & 
 & &  N &  N_{\mathrm{gauge}} &  N_{\mathrm{phys}} \\
\hline \\[-2ex]
\mathbf{1}&0 & (0,0)&
 && 1 & 0 & 1\\[0.5ex]
\mathbf{2}&0 &(\frac12,\frac12)&
 &  & 2 & 1 & 1\\[1.5ex]
\mathbf{3}_{j} & 0 &(j-\frac12,j-\frac12)&  \frac32\le j  &
  & 3 &  1 & 2\\[1.5ex]
\mathbf{2}_{j} &1 &(j,j-1)&   \frac32 \le j &
& 2 &1 & 1 \\[0.5ex]
&-1 &(-j-1, -j)&   j\le -\frac32&
& 2 &1 & 1
\\[1.5ex]
\mathbf{1}_{j} & 2&(j+\frac12,j-\frac32) & \frac32 \le j&
& 1 &0 & 1\\[0.5ex]
 & -2&(-j-\frac32,-j+\frac12) & j\le -\frac32 &
& 1 &0 & 1
\end{array}
$$
\caption{The irreducible representations with $N_{\mathrm{phys}}>0$.}
\label{table-irreps}
\end{table}

\section{Five ODEs}
\label{sect:fiveodes}

The fluctuation $B_{i}^{j}(\hat x)$ breaks up into
a sum of degrees of freedom 
in the irreducible representations.
\eq
\label{eq:qjLjR}
B = \sum_{j_{L},j_{R}} C_{j_{L},j_{R}}\,q_{j_{L},j_{R}}
\qquad
q_{j_{L},j_{R}} \in \Complexes^{N(j_{L},j_{R})}
\otimes (j_{L},j_{R})
\en
The normalization constants $C_{j_{L},j_{R}}$ are chosen so that
\eq
\frac1 {\epsilon g ^{2}} \int -\tr \left(
B^{i} B_{i}\right)
\, \sqrt{-\hat g} \,d^{3}\hat x
= \sum_{j_{L},j_{R}} \frac12 q_{j_{L},j_{R}}^{t} q_{j_{L},j_{R}}
\en
Each $q_{j_{L},j_{R}}$
is an independent degrees of freedom
in the quadratic action (\ref{eq:quadraticYMaction})
governed by an action
\eqg
\label{eq:qaction}
\frac1\hbar S = \int dz \;  \left(
- \frac12 \frac{d q^{t}}{dz} \frac{d q}{dz}
+ \frac12  q(z)^{t} \bK(z) q(z) \right)
\qquad
\bK(z)^{t} = \bK(z)
\eng
where $\bK(z)$ is an $N{\times}N$ 
matrix acting on the factor $\Complexes^{N}$ in (\ref{eq:qjLjR}).
$\bK(z)$ is a symmetric matrix because the linear operator 
$\bK(z)$ in the quadratic Yang-Mills action (\ref{eq:quadraticYMaction})
is symmetric.
The equation of motion is the ode
\eq
\label{eq:eqnofmotion}
\frac{d^{2}q}{dz^{2}} + \bK(z) q(z) = 0
\en
When $N_{\gauge}=1$ there will be a nonzero solution
\eq
\label{eq:eqnofmotiongauge}
\left(\frac{d^{2}}{dz^{2}} + \bK(z) \right) w_{\gauge}(z) = 0
\qquad
w_{\gauge}(z) \in \Complexes^{N}
\en
The matrix $\bK(z)$ and gauge solution $w_{\gauge}(z)$
are shown below for the five sets of representations
in Table~\ref{table-irreps}.
Their calculation is shown in \cite{Friedan2022:StabilitySuppMat}.
The method is:
\begin{enumerate}
\item The expansion of the triple tensor product in (\ref{eq:Bdecomp}) is written
\eqg
j_{1}\otimes j_{2}\otimes j_{3}
= \mathop\oplus\limits_{J} \Complexes^{N(J)}\otimes J
\qquad
j_{1}=j_{2}=1
\quad
j_{3}=j_{R}
\eng
The $\Spin(4)$-invariant operators $\starnabla$ and $\Gamma$ defined 
in (\ref{eq:Gammastarnabla})
are expressed as
\eq
\Gamma = C_{12} -2
\qquad
\starnabla-\Gamma = 2 C_{23} - j_{3}(j_{3}+1)
\en
where $C_{12}$ is the Casimir operator on the factor $j_{1}\otimes j_{2}$,
$C_{23}$ the Casimir on $j_{2}\otimes j_{3}$,
and $C_{3}$ the Casimir on $j_{3}$.

\item
$C_{12}$ and $C _{23}$ act as matrices on $\Complexes^{N(J)}$
but they do not commute.
Each can be diagonlized in a canonical basis,
but not simultaneously.
The unitary matrix $U$ that translates
between the two diagonalizations is given by
Wigner 6-j symbols or Racah W-coefficients.

\item
The Casimir eigenvalues and the matrix $U$ are combined
to find the matrices $\bK(z)$.

\item The gauge solution $w_{\gauge}(z)$
is the unique solution of 
(\ref{eq:eqnofmotiongauge}) linear in $b(z)$.
\end{enumerate}

The local physics in the CGF cosmology is expressed in terms of the 
scale $a=\epsilon R$.
The physical length scale of a fluctuation in the representation 
$(j_{L},j_{R})$ is $a/p$ where
\eq
p = 2  (C_{j_{L}}+C_{j_{R}})^{\frac12}\epsilon
\en
The eigenvalue of the dimensionful physical laplacian on fluctuations is
\eq
\frac{p^{2}}{a^{2}} = \frac{ 4  (C_{j_{L}}+C_{j_{R}})}{R^{2}}
\en
The numerical data
for the series $\mathrm{3}_{j}$, $\mathrm{2}_{j}$, $\mathrm{1}_{j}$
is parametrized by $\epsilon,\,p$
instead of $\epsilon,\,j$.

\newpage

\noindent
{\bf ode 1}
\qquad $N=1$, \; $N_{\gauge}=0$
\qquad
$(j_{L},j_{R}) = (0,\,0)$

\eqg
\label{eq:ode1}
\bK(z) = 6 b(z)^{2} -2\epsilonb ^{2}
\eng
\vskip4ex

\noindent
{\bf ode 2}
\qquad $N=2$, \; $N_{\gauge}=1$
\qquad
$(j_{L},j_{R})  =({\textstyle\frac12,\,\frac12})$
\qquad
$\sigma = \frac{\epsilonb }{\sqrt2} $
\vskip1ex

\eqg
\label{eq:ode2}
\bK(z)=
\begin{pmatrix}
6& 0\\[1ex]
0& 2
\end{pmatrix}
b(z)^{2}
+
\begin{pmatrix}
0& -6\\[1ex]
-6& 0
\end{pmatrix}
\sigma  b(z)
+
\begin{pmatrix}
0& 0\\[1ex]
0& 1
\end{pmatrix}
2\sigma^{2}
\\[3ex]
w_{\gauge}(z)
=
\begin{pmatrix}
0\\[1ex]
1 
\end{pmatrix}
b(z)
+\begin{pmatrix}
1\\[1ex]
0
\end{pmatrix}
\sigma
\eng
\vskip4ex

\noindent
{\bf ode$\;\mathbf{3}_{j}$}
\qquad $N=3$, \; $N_{\gauge}=1$
\qquad
$(j_{L},j_{R}) 
= (j-{\textstyle\frac12},\,j-{\textstyle\frac12})\qquad
j \ge \frac32$
\vskip1ex

\eqg
\label{eq:ode3j}
p = \epsilon\sqrt{4j^{2}-1}
\qquad
\sigma = \epsilonb 
\sqrt\frac{2}{3}\sqrt{j^{2}-\frac1{4}}
=\frac{p}{\sqrt6}
\qquad
\alpha =  \sqrt2 \sqrt{\frac{j^{2}-1}{j^{2}-\frac1{4}}}
\\[3ex]
\bK(z)=
\begin{pmatrix}
6&0& 0\\[1ex]
0&2& 0\\[1ex]
0&0& 0
\end{pmatrix}
 b(z)^{2}
+
\begin{pmatrix}
0 & -6 & 0\\[1ex]
-6 & 0 & 0\\[1ex]
0 & 0 & 0
\end{pmatrix}
\sigma   b(z)
+  
\begin{pmatrix}
\alpha ^{2} & 0 &\alpha \\[1ex]
0 & \alpha ^{2}+1 & 0\\[1ex]
\alpha & 0 & 1
\end{pmatrix}
2 \sigma^{2}
\\[2ex]
w_{\gauge}(z)
=
\begin{pmatrix}
0 \\[1ex]
1\\[1ex]
0
\end{pmatrix}
b(z)
+
\begin{pmatrix}
1 \\[1ex]
0\\[1ex]
-\alpha 
\end{pmatrix}
\sigma 
\eng
\vskip4ex

\noindent
{\bf ode$\;\mathbf{2}_{j}$}
\qquad $N=2$, \; $N_{\gauge}=1$
\qquad
$(j_{L},j_{R}) = 
\left\{
\begin{array}{l@{\qquad}l}
\big(j,\, j-1\big)& j \ge \frac32\\[1.5ex]
\big(-j-1,\, -j\big)& j \le - \frac32
\end{array}
\right.$
\vskip3ex

\eqg
\label{eq:ode2j}
p = 2j\epsilon
\qquad
\sigma = j \epsilonb =\frac{p}{2}
\qquad
\alpha=
\sqrt{1- \frac{1}{j^{2}}}
\\[3ex]
\bK(z) = 
\begin{pmatrix}
2& 0\\[1ex]
0& 0
\end{pmatrix}
b(z)^{2}
+
\begin{pmatrix}
-2& 0\\[1ex]
0 & 2
\end{pmatrix}
 \sigma b(z)
+
\begin{pmatrix}
\alpha^{2}& \alpha\\[1ex]
\alpha & 1
\end{pmatrix}
2  \sigma^{2}
\\[3ex]
w_{\gauge}(z)
=
\begin{pmatrix}
1\\[1ex]
0
\end{pmatrix}
b(z)
+
\begin{pmatrix}
1\\[1ex]
- \alpha
\end{pmatrix}
\sigma
\eng
\newpage
\noindent
{\bf ode$\;\mathbf{1}_{j}$}
\qquad $N=1$, \; $N_{\gauge}=0$
\qquad
$(j_{L},j_{R}) = 
\left\{
\begin{array}{l@{\qquad}l}
\big(j+{\textstyle\frac12},\, j-{\textstyle\frac32}\big)& 
j \ge \frac32
\\[1.5ex]
\big(-j-\frac32,\, -j+\frac12\big)& 
j \le -\frac32
\end{array}
\right.$
\vskip3ex

\eqg
\label{eq:ode1j}
p = \epsilon\sqrt{4j^{2}+3}
\qquad
\sigma = j \epsilonb =\frac{\sqrt{p^{2}-3\epsilon^{2}}}{2}
\qquad
\alpha = \sqrt{1 + \frac1{4j^{2}}}
\\[3ex]
\bK(z) = 
4
\sigma b(z) + 
4 \alpha^{2}
\sigma^{2}
\eng

\section{Time-translation zero-mode}

First consider
the $\Spin(4)$-invariant sector
$(j_{L},j_{R}) = (0,0)$.
The $(0,0)$ perturbation is
governed by {\bf ode 1}.
The infinitesimal time translation is a 
zero-mode,
\eq
\left(\frac{d^{2}}{dz^{2}} + \bK(z)\right) b'(z)
= 0
\en
Time translation is an exact symmetry of the Yang-Mills theory,
so the path integral over the  $(0,0)$ perturbations
must extend to a path integral over all the time 
translations of $b(z)$.  
These form a circle --- the periodic imaginary time trajectory.

Change variable,
\eq
\tilde q(z) = \frac{q(z)}{b'(z)}
\en
The equation of motion and action become
\eqg
\frac1{b'(z) } \frac{d}{dz} \left(b'(z) ^{2} \frac{d \tilde 
q}{dz}\right)
= 0
\qquad
S = \int dz\;  \left(- b'(z)^{2}  \left( 
\frac{d \tilde 
q}{dz}
\right) ^{2}\right)
\eng
Now change time variable,
\eq
\frac{d\tilde z}{dz} = \frac{1}{b'(z) ^{2}}
\en
The equation of motion and action become
\eqg
\label{eq:freeparticle}
\frac{d^{2} \tilde q}{d \tilde z^{2}} =0
\qquad
S = \int d\tilde z\; \left[-  \left( 
\frac{d \tilde 
q}{d\tilde z}
\right) ^{2}\right]
\eng
The $\Spin(4)$-invariant fluctuations
are equivalent, in the gaussian approximation,
to the fluctuations of an equilibrium
thermal state of a
free particle
moving in a circle.
This is a thermodynamically stable state.

\section{Classical mechanics analytic in complex time $z$}

For the sectors $(j_{L},j_{R}) \ne (0,0)$ transverse to the zero-mode,
a formalism is developed for
the quantum mechanical path
integral for a quadratic hamiltonian
$H(z)$ that is analytic in the time $z$.
The formalism is used to construct the thermal state on 
the fluctuation degrees of freedom $q(z)$.
Thermodynamic stability implies a certain 
``property {\bf P}'' of the matrix-valued function $\bK(z)$.
Conversely,
when property {\bf P} is satisfied $q(z)$
becomes canonically equivalent to an ordinary harmonic oscillator at finite 
temperature,
which is thermodynamically stable.
Property {\bf P} is equivalent to thermodynamically stability.
Property {\bf P} is then verified numerically
for each of the four remaining ODEs.

\subsection{First-order phase-space formalism}

Introduce the phase-space degree of freedom
\eq
\cQ = \begin{pmatrix}
q\\
p
\end{pmatrix}
\en
a rank $2N$ block vector.  $p$ is the momentum conjugate to $q$.  The classical equation of motion
becomes the first-order differential equation
\eqg
\label{eq:equationofmotion}
\frac{d\cQ}{dz} + \cA(z) \cQ(z) = 0
\qquad
\cA(z) =  \begin{pmatrix}
0 & - 1 \\[1ex]
\bK(z) & 0
\end{pmatrix}
\eng
i.e.,
\eq
\frac{dq}{dz} = p(z)
\qquad
\frac{dp}{dz} = -\bK(z) q(z)
\en
The first-order differential equation
(\ref{eq:equationofmotion}) 
has locally analytic solutions 
away from the poles of $\bK(z)$.
Global solutions are multi-valued in $z$ because of monodromy around the poles.

\subsection{Path-dependent classical propagator}

Let $C$ be a path in the complex $z$ plane
avoiding the poles.
For points  $z_{1},\,z_{2}$ on $C$,
the path-dependent classical propagator
$\cP_{C}(z_{2},z_{1})$ is the integral of
\eq
\left(\frac{d  }{dz} + \cA(z)\right) \cP_{C}(z,z_{1}) = 0
\qquad
\cP_{C}(z_{1},z_{1}) = 1
\en
along the path from $z_{1}$ to $z_{2}$.
The propagator
$\cP_{C}(z_{2},z_{1})$ is a $2N{\times}2N$ complex matrix
that depends only on the homotopy class of the path from 
$z_{1}$ to $z_{2}$.
For any three points $z_{1},\,z_{2},\,z_{3}$ on $C$,
\eq
\cP_{C}(z_{3},z_{1}) = \cP_{C}(z_{3},z_{2}) 
\cP_{C}(z_{2},z_{1})
\qquad
\cP_{C}(z_{2},z_{1})^{-1} = \cP_{C}(z_{1},z_{2})
\en
$\bK(z)=\bK(z)^{t}$ implies
\eq
\cA(z)^{t} \Omega +\Omega \cA(z) = 0
\qquad
\Omega = i \begin{pmatrix}
0 & 1 \\[1ex]
-1 & 0
\end{pmatrix}
\en
so the propagator is a complex symplectic matrix
\eq
\label{eq:symplectic}
\cP_{C}(z_{2},z_{1})^{t} \,\Omega\, \cP_{C}(z_{2},z_{1}) = \Omega
\qquad
\cP_{C}(z_{2},z_{1}) \in \mathrm{Sp}(2N,\Complexes)
\en
The solution of the equation of motion 
(\ref{eq:equationofmotion}) along $C$ is
\eq
\cQ(z) =  \cP_{C}(z,z_{1}) \cQ(z_{1})
\en

\subsection{Gauge symmetry}
When there is a gauge solution $w_{\gauge}(z)\in \Complexes^{N}$ then
\eq
\cW_{\gauge}(z) =
\begin{pmatrix}
w_{\gauge}(z) \\[1ex]
w_{\gauge}'(z)
\end{pmatrix}
\en
is a solution of the first-order equation of motion (\ref{eq:equationofmotion}).
The $\Omega$-complement $\cW_{\gauge}(z)^{\perp}$ is
the subspace of $\Complexes^{2N}$
\eq
\cW_{\gauge}(z)^{\perp} = \left\{
\cW \in \Complexes^{2N} \colon 
\cW_{\gauge}(z)^{t} \Omega \cW = 0
\right\}
\en
$\Omega$ is antisymmetric so
\eq
\cW_{\gauge}(z) \in \cW_{\gauge}(z)^{\perp}
\en
The physical phase-space at time $z$ is the quotient space
\eq
\cV_{\phys}(z) =\cW_{\gauge}(z)^{\perp}/\Complexes \cW_{\gauge}(z)
\en
The physical degrees of freedom live in $\cV_{\phys}(z) \otimes 
(j_{L},j_{R})$.
The propagator  preserves 
$\cW_{\gauge}(z)$ and it preserves $\Omega$,
\eq
\cP_{C}(z_{2},z_{1})^{t} \,\Omega\, \cP_{C}(z_{2},z_{1}) = \Omega
\qquad
\cW_{\gauge}(z_{2}) = \cP_{C}(z_{2},z_{1})\cW_{\gauge}(z_{1}) 
\en
so the propagator acts as a linear map from $\cV_{\phys}(z_{1})$ to 
$\cV_{\phys}(z_{2})$,
\eq
\cP_{C}(z_{2},z_{1})\colon \cV_{\phys}(z_{1}) \rightarrow \cV_{\phys}(z_{2})
\en

\section{Quantum mechanics analytic in complex time $z$}


Quantization makes $q$ and $p$  operators on Hilbert space, i.e. 
rank $N$ vectors whose entries are operators.
The canonical commutation relations
\eqg
\left[ p_{b}, \,q^{a}\right]
= i\, \delta^{a}_{b}
\qquad
[q_{b},\,q^{a}] = 0
\qquad
[p_{b},\,p^{a}] = 0
\eng
are expressed by the matrix equation
\eq
\label{eq:ccr}
\left( \cQ \cQ^{t}\right)^{t}- \cQ \cQ^{t}  = \Omega
\en
where the matrix transpose does not change the operator ordering.
For example,
\eqg
\left(p\, q^{t}\right)^{t}{}^{a}_{b} -\left( q\,p^{t}\right)^{a}_{b}
=p_{b} \,q^{a} - q^{a} \, p_{b}
= \left[ p_{b}, \,q^{a}\right]
= i\, \delta^{a}_{b}
\eng

\subsection{Path-dependent time evolution}

The hamiltonian depends analytically on the complex time $z$.
\eq
H(z) = \frac12 p^{t}p + \frac 12 q^{t} \bK(z) q
\en
The time evolution of
the state vector $\psi(z)$ is
given by the Schr\"odinger equation.
\eq
\label{eq:Schrodinger}
\frac{d\psi}{dz}  = i H(z) \psi(z) 
\en
Again, for $z_{1},\,z_{2}$ two points on a path $C$ in the complex 
$z$ plane avoiding the poles,
the path-dependent time evolution operator $U_{C}(z_{2},z_{1})$ is constructed
by integrating
\eq
\label{eq:timeevolutionoperator}
\left(\frac{d}{dz} - i H(z)\right) U_{C}(z,z_{1}) =0
\qquad
U_{C}(z_{1},z_{1}) = 1
\en
along the path from $z_{1}$ to $z_{2}$.
Again, $U_{C}(z_{2},z_{1})$ depends only on the homotopy class of 
the path from $z_{1}$ to $z_{2}$.
For any three points $z_{1},\,z_{2},\,z_{3}$ on $C$,
\eq
U_{C}(z_{3},z_{1}) = U_{C}(z_{3},z_{2}) U_{C}(z_{2},z_{1})
\qquad
U_{C}(z_{2},z_{1})^{-1} = U_{C}(z_{1},z_{2})
\en
The solution $\psi(z)$ of the Schr\"odinger equation along $C$ is
\eq
\psi(z) = U_{C}(z,z_{1}) \psi(z_{1})
\en
If $H(z)$ were constant then $U_{C}(z_{2},z_{1})$ would be 
independent of the path
\eq
U_{C}(z_{2},z_{1}) = e^{i(z_{2}-z_{1})H}
\en

\subsection{Path-dependent path integral}

For $C$  a path from $z_{1}$ to $z_{2}$
the time evolution operator along $C$
is given by the phase-space path integral
\eqg
U_{C}(z_{2},z_{1}) = \int \cD \cQ \; 
e^{i S_{C}[\cQ]/\hbar}
\eng
using the phase-space action
\eqg
\frac1{\hbar} S_{C}[\cQ] = \int_{C} dz\; \frac1{2i}\cQ^{t} \Omega \left( \frac{d}{dz} + \cA \right)\cQ
=  \int_{C} dz\;  \left(
-p(z)^{t}\frac{dq}{dz}
+H(z)
\right)
\eng
Integrating out $p(z)$ gives the ordinary path-integral
with the action (\ref{eq:qaction})
\eqg
\int \cD p \; e^{iS_{C}[\cQ]/\hbar}
=
e^{iS_{C}[q]/\hbar}
\\[1ex]
\frac1{\hbar}  S_{C}[q] = \int_{C} dz \;  \left(
- \frac12 \frac{d q^{t}}{dz} \frac{d q}{dz}
+ \frac12  q(z)^{t} \bK(z) q(z) \right)
\eng

\subsection{Operator insertions}

Suppose $C$ is a path from $z_{1}$ to $z_{2}$.
For $z$ in $C$,
the insertion of $\cQ(z)$ in the path integral is
\eq
\int \cD \cQ \; e^{iS_{C}[\cQ]/\hbar} \; \cQ(z)
= U_{C}(z_{2},z) \cQ\, U_{C}(z,z_{1})
\en
The Schwinger-Dyson equation of the path integral
is the equation of motion
\eq
\left( \frac{d}{dz} + \cA(z) \right)
\int \cD \cQ \; e^{iS_{C}[\cQ]/\hbar} \; \cQ(z)
= 0
\en
Equivalently,
\eqa
\left( \frac{d}{dz} + \cA(z) \right)  \big[U_{C}(z_{2},z) & \cQ\, 
 U_{C}( z,z_{1})\big]
=
\\[1ex]
&=U_{C}(z_{2},z) \big( - [iH(z),\,\cQ]+\cA(z)\cQ\big) U_{C}(z,z_{1})
\\[1ex]
&= 0
\ena
Integrating the equation of motion along $C$,
\eq
\int \cD \cQ \; e^{iS_{C}[\cQ]/\hbar} \; \cQ(z')
=
\cP_{C}(z',z)
\int \cD \cQ \; e^{iS_{C}[\cQ]/\hbar} \; \cQ(z)
\en

\section{Imaginary time path integral}

Choose a real time $t$  not in the set $2K \Integers$,
i.e.~such that $2mK < t < 2(m+1)K$ for some integer $m$.
Let $C_{t}$ be the straight vertical path through $t$.
$C_{t}$ avoids the poles since $t\ne 2mK$.
Let $C'_{t}$ be the path from $t$ to $t+4K'i$.
The expectation values in the thermal state are constructed in two steps.
\begin{enumerate}
\item
Perform the gaussian path integral on the path $C'_{t}$ with 
periodic boundary conditions, $\cQ(t+4K'i) = \cQ(t)$.
\item
Then integrate over the 
time-translation zero-mode.
\end{enumerate}
The result will be independent of the choice of
$t$ (shown in section~\ref{sec:thermalstateindoft} below).
Suppose the first step produces a stable state.
Then the underlying time translation symmetry
implies that all the time translated gaussian integrals
are also stable.
The second step is then an integral of stable gaussian integrals, so 
the resulting thermal state is stable.
Stability of the thermal state follows
from stability of the gaussian integral produced in step 1.

The two-point expectation values
determine all the gaussian expectation values.
The gaussian two-point expectation values are expressed by the matrix
\eqg
\cG_{t}(z_{2},z_{1}) =\expval{\cQ(z_{2})\,\cQ(z_{1})^{t}}_{t}
\\[1ex]
z_{1} = t+\tau_{1}i
\qquad
z_{2} = t+\tau_{2}i
\qquad
0\le\tau_{1}\le\tau_{2}\le 4K'
\eng
The expectation value $\expval{\cdot}_{t}$ is
given by the path integral with periodic boundary 
conditions  $\cQ(t+4K'i)=\cQ(t)$
which gives the operator trace.
\eqa
\expval{\cQ(z_{2})\,\cQ(z_{1})^{t}}_{t}&=  \frac1{Z} 
\int \cD \cQ \; e^{iS_{C'_{t}}[\cQ]/\hbar} \; 
\cQ(z_{2})\; \cQ(z_{1})^{t}
\\[1ex]
&=
\frac1{Z_{t}} \,\tr \; \left[U_{C_{t}}(t+4K'i,z_{2})\,\cQ\, 
U_{C_{t}}(z_{2},z_{1}) \,\cQ^{t}\, U_{C_{t}}(z_{1},t)
\right]
\ena
$Z_{t}$ is the normalizing constant
\eq
Z_{t} = \int \cD \cQ \; e^{iS_{C'_{t}}[\cQ]/\hbar}
= \tr \;U_{C_{t}}(t+4K'i,t)
\en 
The expectation values obey the equation of motion in both variables,
\eq
\label{eq:cGeqnofmotion}
\cG_{t}(z_{2},z_{1}) = \cP_{C_{t}}(z_{2},t) \cG_{t}(t,t)  \cP_{C_{t}}(z_{1},t)^{t}
\en
so the gaussian expectation values are all determined by one
matrix, $\cG(t,t)$.

Periodicity is the condition
\eq
\cG_{t}(t+4K'i,z_{1}) = \cG_{t}(z_{1},t)^{t}
\en
By (\ref{eq:cGeqnofmotion}) this is
\eqa
\label{eq:cGperiodic}
\Moni (t)\, \cG_{t}(t,t) &= \cG_{t}(t,t)^{t}
\qquad
\Moni (t) = \cP_{C}(t+4K'i, t)
\ena
$\Moni (t)$ is the imaginary period monodromy matrix at time $t$.

The canonical commutation relations (\ref{eq:ccr}) imply
\eq
\label{eq:cGccr}
\cG_{t}(t,t)^{t}-\cG_{t}(t,t) = \Omega
\en
Combining (\ref{eq:cGperiodic}) and (\ref{eq:cGccr}) gives
\eq
\left(\Moni (t)-1\right) \cG_{t}(t,t)  = \Omega
\en
Define
\eq
\label{eq:Mtt}
\cH(t) = \Omega \left(\Moni (t)-1\right)
\en
$\cH(t)$ is a well defined quadratic form on the physical phase space
$\cV_{\phys}(t)$ because, if there is a gauge solution
$\cW_{\gauge}(z)$, then it has the same periodicities as $b(z)$.
In particular, $\cW_{\gauge}(t+4K'i)=\cW_{\gauge}(t)$ so
$\left(\Moni (t) -1\right) \cW_{\gauge}(t)=0$.
If $\cH(t)$ is invertible on $\cV_{\phys}(t)$,
\eq
\cG_{t}(t,t) = \cH(t)^{-1}
\en
If $\cH(t)$ fails to be invertible on $\cV_{\phys}(t)$, then there is
a solution of the equation of motion that is periodic in imaginary 
time modulo gauge symmetry.  This would be an accidental physical 
zero-mode.  It would be necessary to go beyond the gaussian 
approximation to test for stability.

Here the strong meaning of stability is taken, stability in the 
gaussian approximation,
allowing neither instabilities nor accidental zero modes.

\section{Complex conjugation and operator adjoints}
Use the following notation for complex conjugation and 
the operator adjoint:\\[1ex]
\hspace*{3em}
\begin{tabular}{c@{\;\;=\;\;}l}
$\bar z$ & the complex conjugate of a complex number $z$\\[1ex]
$\overline{O}$ & the adjoint of an operator $O$\\[1ex]
$M^{\dagger}=\overline{M}^{t}$ & the adjoint of a matrix $M$ of operators 
or complex numbers
\end{tabular}
\vskip2ex

\noindent
$b(z)$ is real on the real axis, $\overline{b(z)} = b(\bar z)$,
so $\bK(z)=\bK(z)^{t}$ is real on the real axis
\eq
\overline{\bK(z)} = \bK(\bar z) = \bK(z)^{\dagger} 
\en
The representations $(j_{L},j_{R})$ in the 
decomposition are all real, $j_{L}-j_{R}\in \Integers$,
so
\eq
\overline{q} = q
\qquad
\overline{p} = q
\qquad
\overline{\cQ} = \cQ
\qquad
\cQ^{\dagger} = \cQ^{t}
\en
so the hamiltonian is real (self-adjoint) on the real axis
\eq
\overline{H(z)} = H(\bar z)  = H(z)^{\dagger}
\en
So 
the time evolution operator given by (\ref{eq:timeevolutionoperator}) 
satisfies, for any path $C$,
\eq
\overline{U_{C}(z_{2},z_{1})} = U_{C}(z_{2},z_{1})^{\dagger} = 
U_{\overline{C}}(\bar{z_{2}}, \bar{z_{1}})^{-1}
=U_{\overline{C}}(\bar{z_{1}}, \bar{z_{2}})
\en
$\cA(z)$ is real on the real axis because $\bK(z)$ is,
\eq
\overline{\cA(z)} = \cA(\bar z)
\en
Therefore, for any path $C$,
\eq
\overline{\cP_{C}(z_{2},z_{1})} = 
\cP_{\overline{C}}(\bar{z_{2}}, \bar{z_{1}})
\en
In particular, for $C_{t}$ the vertical path through $t$,
\eq
\overline{\cP_{C_{t}}(z_{2},z_{1})} = \cP_{\overline{C_{t}}}(\bar{z_{2}}, \bar{z_{1}})
= \cP_{C_{t}}(\bar{z_{2}}, \bar{z_{1}})
\en
Imaginary time periodicity implies
\eq
\cP_{C_{t}}(z_{2},z_{1}) =\cP_{C_{t}}(z_{2}+4K'i,z_{1}+4K'i)
\en
so the imaginary period monodromy matrix satisfies
\eqg
\overline{\Moni (t) } = \overline{\cP_{C_{t}}(t+4K'i,t) } = 
\cP_{C_{t}}(t-4K'i,t) 
= \cP(t,t+4K'i)  = \Moni (t)^{-1}
\eng
Then, by the symplectic property (\ref{eq:symplectic}),
\eq
\label{eq:cMdagger}
\Moni (t)^{\dagger}\,\Omega = \Omega\, \Moni (t)
\en
It follows that $\cH(t)$ defined in (\ref{eq:Mtt}) is hermitian.
\eq
\label{eq:hermiticity}
\cH(t)^{\dagger} = (\Moni (t)^{\dagger}-1) \Omega
= (\Omega\Moni (t) -\Omega) 
=\cH(t)
\en

\section{Property {\bf P}}

\subsection{Heisenberg picture}

Write $\Reals$ for the path along the real time axis.
The real time evolution operator is unitary
\eq
U_{\Reals}(t_{2},t_{1})^{\dagger} = U_{\Reals}(t_{2},t_{1})^{-1} = 
U_{\Reals}(t_{1},t_{2})
\en
The Heisenberg picture operator $\cQ(t)$ is 
determined by
\eq
\psi_{2}^{\dagger}(t) \,\cQ \psi_{1}(t) 
= \psi_{2}^{\dagger}(0) \,\cQ(t) \psi_{1}(0) 
\en
for arbitrary states $\psi_{1,2}(0)$.
So
\eq
\cQ(t) = U_{\Reals}(t,0)^{\dagger} \cQ U_{\Reals}(t,0)
\en
and
\eq
\cQ(t)^{\dagger} =  U_{\Reals}(t,0)^{\dagger} \cQ^{t} U_{\Reals}(t,0) 
= \cQ(t)^{t}
\en

\subsection{Stability requires Property {\bf P}}

The matrix of operators
\eq
\cQ(t) \,\cQ(t)^{t}= \cQ(t) \,\cQ(t)^{\dagger}
\en
is hermitian and positive so the expectation value in a stable state 
must be hermitian and positive.
\eq
\cG_{t}(t,t) = \expval{\cQ(t) \,\cQ(t)^{\dagger}}_{t} = \cG_{t}(t,t)^{\dagger} 
\qquad
\cG_{t}(t,t) > 0
\en
so $\cG_{t}(t,t) ^{-1}=\cH(t)$ must be hermitian and positive
\eq
\cH(t)^{\dagger} = \cH(t)
\qquad \cH(t) > 0
\en
$\cH(t)$ is given by equation (\ref{eq:Mtt}).
It has already been shown to be hermitian, equation (\ref{eq:hermiticity}).
Stability requires
\eq
\text{\bf Property {\bf P}}
\qquad\quad
\cH(t)=\Omega (\Moni (t) - 1) > 0
\en
as hermitian form on the physical phase space $\cV_{\phys}(t)$.

\section{Path-independent time evolution in region $R_{m}$}

\begin{figure}
\begin{center}
\begin{tikzpicture}[baseline=2ex,x=1.05cm,y=0.7cm]
%
\foreach \y in {-3,-1,1,3}
{
\node at (-4,\y) {$\scriptscriptstyle \otimes$};
\node at (-2,\y) {$\scriptscriptstyle \otimes$};
\node at (0,\y) {$\scriptscriptstyle \otimes$};
\node at (2,\y) {$\scriptscriptstyle \otimes$};
\node at (4,\y) {$\scriptscriptstyle \otimes$};
\node at (6,\y) {$\scriptscriptstyle \otimes$};
\draw[dashed,thick] (6.1,\y) -- (6.5,\y);
\draw[dashed,thick] (4.1,\y) -- (5.93,\y);
\draw[dashed,thick] (2.1,\y) -- (3.93,\y);
\draw[dashed,thick] (-0.1,\y) -- (-1.93,\y);
\draw[dashed,thick] (-2.1,\y) -- (-3.93,\y);
\draw[dashed,thick] (-3.1,\y) -- (-4.5,\y);
}
\node at (0,-1-.4){$\scriptstyle 2mK-K'i$};
\node at (0,1-.4){$\scriptstyle 2mK+K'i$};
\node at (2,-1-.4){$\scriptstyle 2(m+1)K-K'i$};
\node at (2,1-.4){$\scriptstyle 2(m+1)K+K'i$};
\draw[dotted,thick] (4,3.7) -- (4,4.2);
\draw[dotted,thick] (4,-3.7) -- (4,-4.2);
\draw[dotted,thick] (-2,3.7) -- (-2,4.2);
\draw[dotted,thick] (-2,-3.7) -- (-2,-4.2);
%
%
\end{tikzpicture}
\end{center}
\caption{The simply connected open region $R_{m}$ is formed by cutting from the complex plane
the indicated set of half-infinite horizontal lines containing the poles.
}
\label{fig:polesandRm}
\end{figure}
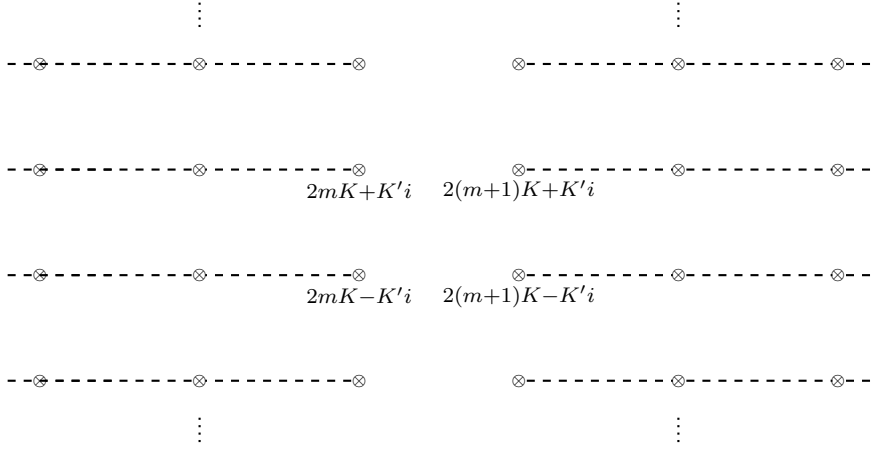

For each integer $m$
form the region $R_{m} \subset \Complexes$
shown in Figure~\ref{fig:polesandRm},
cutting out from the complex plane 
a set of half-infinite horizontal lines 
containing all the poles,
\eqa
R_{m} = \Complexes &- \mathop\cup_{n\in\Integers}
\left\{t+ (2n+1) K' i\colon
 (2m+1)K \le t
\right\}
\\[1ex]
&- \mathop\cup_{n\in\Integers}
\left\{t+ (2n+1) K' i\colon
t \le 2m K
\right\}
\ena
$R_{m}$ is simply connected so 
time evolution within $R_{m}$ is path-independent:
\eq
\begin{aligned}
U_{m}(z_{2},z_{1}) &= U_{C}(z_{2},z_{1})
\\[1ex]
\cP_{m}(z_{2},z_{1}) &= \cP_{C}(z_{2},z_{1})
\end{aligned}
\qquad
z_{1},\,z_{2} \in R_{m}
\quad
C \subset R_{m}
\en
Solutions of the equation of motion are single-valued in 
$R_{m}$ but are discontinuous across the cuts.
The path-independent propagator in $R_{m}$ satisfies
\eqg
\cP_{m}(z,z) = 1
\qquad
\cP_{m}(z_{2},z_{1})^{-1}=\cP_{m}(z_{1},z_{2})
\\[1ex]
\cP_{m}(z_{3},z_{2})\cP_{m}(z_{2},z_{1}) = \cP_{m}(z_{3},z_{1})
\eng
Translation by $4K'i$ takes $R_{m}$ to $R_{m}$.
Translation by $2K+2K'i$ takes $R_{m}$ to $R_{m+1}$.
Translation by $4K$ takes $R_{m}$ to $R_{m+2}$.
So the periodicities imply
\eqa
\cP_{m}(z_{2},z_{1})
&=\cP_{m}(z_{2}+4K'i,z_{1}+4K'i) 
\\[2ex]
&= \cP_{m+1}(z_{2}+2K+2K'i,z_{1}+2K+2K'i)
\\[2ex]
&=
\cP_{m+2}(z_{2}+4K,z_{1}+4K) 
\ena
$\overline{R_{m}}=R_{m}$ so
\eq
\overline{\cP_{m}(z_{2},z_{1})}
= \cP_{m}(\bar{z_{2}},\bar{z_{1}})
\en

\section{Property {\bf P} is independent of $t$}

Now prove property {\bf P} is independent of $t$.
That is,  prove that $\cP(t)$ satisfies property {\bf P} for all $t\in \Reals - 2K 
\Integers$ iff it satisfies property {\bf P} for any $t\in \Reals - 2K 
\Integers$.
So property {\bf P} is a property of the ode, of the matrix function $\bK(z)$.
The proof is in two steps.
Suppose
\eq
2mK < t < 2(m+1)K
\qquad
m\in \Integers
\en

\begin{enumerate}[wide=0pt]
\item If  $t'$ is in the same vertical strip,
$
2mK < t' < 2(m+1)K
$,
then the imaginary period monodromy matrices
\eq
\Moni (t) = \cP_{m}(t+4K'i,t)
\qquad
\Moni (t') = \cP_{m}(t'+4K'i,t')
\en
are related by
\eq
\Moni (t') =  \cP_{m}(t,t')^{-1} \Moni (t)\cP_{m}(t,t')
\en
$\cP_{m}(t,t')$ is a real matrix so
\eqa
\cH(t') &= \Omega (\Moni (t') -1)
= \Omega  \cP_{m}(t,t')^{-1} (\Moni (t) -1)\cP_{m}(t,t')
\\[1ex]
&= \cP_{m}(t,t')^{t} \Omega (\Moni (t) -1)\cP_{m}(t,t')
\\[1ex]
&= \cP_{m}(t,t')^{\dagger} \cH(t) \cP_{m}(t,t')
\ena
Therefore $\cH(t')>0$ iff $\cH(t)>0$.

\item Suppose $t' = t+2K$.  Then
\eqa
\Moni (t) &= \cP_{m}(t+4K'i,t) = \cP_{m}(t+4K'i,t+2K'i) \cP_{m}(t+2K'i,t)
\\[1ex]
&= \cP_{m}(t,t-2K'i) \cP_{m}(t+2K'i,t)
\\[2ex]
\Moni (t') &= \cP_{m+1}(t'+4K'i,t') = \cP_{m}(t+2K+4K'i,t+2K) 
\\[1ex]
&= \cP_{m}(t+2K'i,t-2K'i)
\\[1ex]
&= \cP_{m}(t+2K'i,t) \cP_{m}(t,t-2K'i)
\ena
so
\eqa
\overline{\Moni (t')} -1
&=\cP_{m}(t-2K'i,t)\cP_{m}(t,t+2K'i)-1 
\\[1ex]
&= \cP_{m}(t-2K'i,t)
\left( 1 - \Moni (t)
\right)
\cP_{m}(t,t+2K'i)
\ena
so
\eqa
\overline{\cH(t')}&=\overline{\Omega} \left(\overline{\Moni (t')} 
-1\right)
\\[1ex]
&=  -\Omega \cP_{m}(t-2K'i,t)
\left( 1 - \Moni (t)
\right)
\cP_{m}(t,t+2K'i)
\\[1ex]
&= \cP_{m}(t,t-2K'i)^{t}\Omega \left( \Moni (t)- 1 \right) \cP_{m}(t,t+2K'i)
\\[1ex]
&= \cP_{m}(t,t+2K'i)^{\dagger} \cH(t) \cP_{m}(t,t+2K'i)
\ena
$\cH(t')>0$ iff $\overline{\cH(t')}>0$
so $\cH(t')>0$ iff $\cH(t) >0$.
\end{enumerate}
Together the two steps imply that 
Property {\bf P} is independent of $t$.

\section{Property {\bf P} implies stability}

Now Property {\bf P} is proved to imply existence of
a canonical equivalence between the imaginary time path
integral 
and that of an ordinary 
time-{\it independent} harmonic oscillator,
which is the
finite temperature equilibrium state of the ordinary harmonic 
oscillator, which is manifestly stable.

\subsection{Spectrum of $\Moni (t)$}

Again suppose $t$ is in the vertical strip $2mK < t < 2(m+1)K$.
Leave implicit the dependence on $t$.
\eqg
\cP(z) = \cP_{m}(z,t)
\quad
\Moni  = \Moni (t)=\cP(4K'i)
\quad
\cH = \cH(t)
\quad
\cV_{\phys} = \cV_{\phys}(t)
\eng
Suppose Property {\bf P} is satisfied, so $\cH$ is a 
positive definite hermitian form on $\cV_{\phys}$.
$\Moni $ is self-adjoint with respect to 
the positive definite hermitian form $\cH$
by (\ref{eq:cMdagger}).
\eq
\Moni ^{\dagger} \cH
= \Omega \Moni  \Omega \Omega(\Moni -1)
=\Omega \Moni  (\Moni -1)
=\Omega  (\Moni -1)\Moni 
= \cH \Moni 
\en
Therefore there is a $\cH$-orthonormal basis of $\cV_{\phys}$
consisting of eigenvectors 
$\cW'_{a}$ of $\Moni $ with real eigenvalues $\lambda_{a}$
\eqg
\Moni  \cW'_{a} = \lambda_{a } \cW'_{a}
\qquad
\cW'_{a}{}^{\dagger} \cH \cW'_{b} = \delta_{a,b}
\\[1ex]
\lambda_{1}\ge \lambda_{2} \ge \cdots \ge \lambda_{2 N_{\phys}}
\eng
$\cV_{\phys}$ decomposes into $\cH$-orthogonal eigenspaces $\cV_{\lambda}$
\eq
\cV_{\phys} = \mathop\oplus_{\lambda } \cV_{\lambda} 
\en
None of the eigenvalues $\lambda_{a}$ can equal 1 because $\Moni -1$ 
is invertible on $\cV_{\phys}$.

The complex conjugate of the eigenvalue equation,
\eq
\overline{\Moni } \,\overline{ \cW'_{a} }
=\overline{ \lambda_{a } }\, \overline{ \cW'_{a}}
\en
is
\eq
\qquad
\Moni ^{-1} \overline{\cW_{a}} = \lambda_{a } \, \overline{ \cW'_{a}}
\en
so $\overline{  \cW_{a}} $ is an 
eigenvector with eigenvalue $1/\lambda_{a}$.
Thus
\eq
\overline{\cV_{\lambda}} =  \cV_{1/{\lambda}}
\en
Suppose $\lambda_{a}=\lambda_{b}=\lambda$ so $\cW_{a}$ and $\cW_{b}$ are both in $\cV_{\lambda}$.
Then
\eq
\overline{\cW'_{a}}{}^{t}\Omega \cW'_{b}
= \frac{\overline{\cW'_{a}}{}^{t}\Omega (\Moni -1) \cW'_{b}}{\lambda-1}
= \frac{\overline{\cW'_{a}}{}^{\,t}\cH \cW'_{b}}{\lambda-1}
= \frac{\delta_{ab}}{\lambda-1}
\en
and
\eq
\overline{\cW'_{a}}{}^{t}\Omega \cW'_{b}
= - \cW'_{b}{}^{t}\Omega \overline{\cW'_{a}}
=\frac{-\cW'_{b}{}^{t}\Omega(\Moni -1) 
\overline{\cW'_{a}}}{\frac1{\lambda}-1}
= \frac{\lambda\, \overline{\cW'_{b}}^{\dagger}\cH \overline{\cW'_{a}} }{\lambda-1}
\en
so
\eq
\lambda_{a}=\lambda_{b}
\quad\implies\quad
\overline{\cW'_{b}}^{\dagger}\cH \overline{\cW'_{a}}
=\frac1\lambda_{a} \delta_{ab}
\en
$\cH$ is a positive hermitian form so the last equation implies  all $\lambda_{a} >0$.

If $1/\lambda_{a}\ne\lambda_{b} $ then $\overline{\cW'_{a}}$
is $\cH$-orthogonal to $\cW'_{b}$,
\eq
0 = \overline{\cW_{a}}^{\,\dagger} \cH \cW_{b} =
\cW_{a}^{t} \Omega (\lambda_{b}-1) \cW_{b}
\en
and $\lambda_{b}\ne 1$ so
\eq
\lambda_{a}\lambda_{b}\ne 1 \quad \implies\quad
\cW_{a}^{t} \Omega \cW_{b} = 0
\en
Let
\eqg
\cV_{\phys}^{+} = \mathop\oplus_{\lambda>1 } \cV_{\lambda}
\qquad
\cV_{\phys}^{-} = \overline{\cV_{\phys}^{+}} = \mathop\oplus_{\lambda<1 } \cV_{\lambda}
\\[1ex]
\dim \cV_{\phys}^{+} = \dim \cV_{\phys}^{-} = N_{\phys}
\eng
Define a linear operator $\omega$  on $\cW^{+}_{a}$ 
by letting $a$ range over $\{1,\ldots,N_{\phys}\}$, i.e.~the 
$\lambda_{a}>1$, and letting
\eqg
\cW^{+}_{a} = \sqrt{\lambda_{a}-1} \,\cW'_{a}
\qquad
\cW^{-}_{a} = \overline{\cW^{+}_{a}}
= \sqrt{\lambda_{a}-1} \,\overline{\cW'_{a}}
\\[1ex]
\omega_{a} = \frac1{4K'} \ln \lambda_{a}
\qquad
\omega \cW^{+}_{a} = \omega_{a} \cW^{+}_{a}
\eng
The $\cW^{+}_{a}$ form a basis for $\cV_{\phys}^{+} $
and the $\cW^{-}_{a}$ form a basis for $\cV_{\phys}^{-} $.
The linear operator $\omega$  on $\cW^{+}_{a}$ 
is diagonal with eigenvalues $\omega_{a}$ in this basis.
The operator $\overline{\omega}=\omega$ also acts on $\cW^{-}_{a}$.
In this basis, writing $\beta = 4K'$,
\eqg
\Omega = 
\begin{pmatrix}
0  & -1 \\[1ex]
1 & 0 
\end{pmatrix}
\qquad
\Moni  =
\begin{pmatrix}
e^{\beta \omega} & 0 \\[1ex]
0 & e^{-\beta \omega}
\end{pmatrix}
\qquad
\cH =
\begin{pmatrix}
0  & 1 -  e^{-\beta \omega}\\[1ex]
e^{\beta \omega}-1 & 0 
\end{pmatrix}
\eng
These are identical to the $\Omega$, $\Moni $, and $\cH$ for an ordinary 
time-independent harmonic oscillator with frequencies $\omega_{a}$ in 
equilibrium at inverse temperature $\beta$ in the
phase space basis
of creation 
and destruction  operators
$\cV^{+}_{\phys}$ and $\cV^{-}_{\phys}$.

The spectrum of frequencies $\omega_{a}$ is independent of $t$ by the 
same arguments that gave the $t$ independence of property {\bf P}.

\subsection{Equivalence to a time-independent oscillator}

Property {\bf P} is now used to construct a canonical equivalence
between $\cQ(z)$ and an ordinary
time-independent harmonic oscillator.
Let
\eqg
\cJ = \frac{i  \ln \Moni }{4K' }
\qquad
\cJ \cW^{+}_{a} = i \omega_{a}\cW^{+}_{a}
\qquad
\cJ \cW^{-}_{a} = - i \omega_{a}\cW^{-}_{a} 
\eng
so $\cJ$ is real
\eq
\overline{\cJ\cW_{a}} = \cJ\,\overline{\cW_{a}}
\en
And
\eqa
0  &= \cJ^{\dagger}\cH + \cH \cJ
=  \cJ^{t} \Omega (\Moni -1) + \Omega (\Moni -1)\cJ
\\[1ex]
&=  \left (\cJ^{t} \Omega + \Omega \cJ
\right )  (\Moni -1)
\ena
so $\cJ$ is an infinitesimal symplectic transformation.
\eq
\cJ^{t} \Omega + \Omega \cJ = 0
\en
Define
\eq
\cR(z) = e^{-z \cJ} \cP(z)^{-1}
= \Moni ^{z/4K'i} \;\cP(z)^{-1}
\en
Recall that $\cP(z) = \cP_{m}(z,t)$.
$\cR(z)$ has three essential properties:
\begin{enumerate}
\item $\cR(z)$ is symplectic.
\eq
\cR(z)^{t} \Omega \cR(z) = \Omega
\en
\item $\cR(z)$ is periodic in imaginary time.
\eq
\cR(z+4K'i) = \cR(z)
\en
\item $\cR(z)$ is real on the real axis.
\eq
\overline{\cR(z) } = e^{ - \bar z \overline{\cJ}} \;\overline{\cP(z)}^{-1}
= e^{- \bar z \cJ} \cP(\bar z)^{-1}
= \cR(\bar z)
\en
\end{enumerate}
Make the canonical transformation
\eq
\label{eq:canonicaltransformation}
\cQ_{\HO}(z) =
\begin{pmatrix}
q_{\HO}\\
p_{\HO}
\end{pmatrix}
(z)
= \cR(z) \,\cQ(z)
\en
The first-order equation of motion becomes time-independent.
\eq
\left( \frac{d}{dz} + \cJ\right) \cQ_{\HO}(z)  = 0
\en
The imaginary time phase-space action becomes time-translation 
invariant.
\eqa
\frac1\hbar S_{C'}(\cQ) &= \int_{C'} dz \;  \frac1{2i} \cQ_{\HO}^{t} \Omega 
\left( \frac{d}{dz} + \cJ\right) \cQ_{\HO}
\ena
The real symplectic transformation
\eqg
\begin{pmatrix}
v_{a} \\[1ex]
w_{a}
\end{pmatrix}
=
\sqrt{\frac{e^{\omega_{a}}-1}{2\omega_{a}}}
\begin{pmatrix}
\omega_{a} & \omega_{a} \\[1ex]
i & -i
\end{pmatrix}
\begin{pmatrix}
\cW^{+}_{a}\\[1ex]
\cW^{-}_{a}
\end{pmatrix}
\\[1ex]
\begin{pmatrix}
v_{a} \\[1ex]
w_{a}
\end{pmatrix}
^{t}
\Omega
\begin{pmatrix}
v_{b} \\[1ex]
w_{b}
\end{pmatrix}
=
\begin{pmatrix}
0& i \\[1ex]
-i& 0
\end{pmatrix}
\delta_{a,b}
\eng
brings $\cJ$ to the form
\eq
\label{eq:cJHOform}
\cJ \begin{pmatrix}
v_{a} \\[1ex]
w_{a}
\end{pmatrix}
= 
\begin{pmatrix}
0& -1 \\[1ex]
\omega_{a}^{2}& 0
\end{pmatrix}
\begin{pmatrix}
v_{a} \\[1ex]
w_{a}
\end{pmatrix}
\qquad
\cJ = \begin{pmatrix}
0 & -1 \\[1ex]
\omega^{2} & 0
\end{pmatrix}
\en
So $q_{\HO}$ is an ordinary time-independent harmonic oscillator
satisfying the second-order equation of motion
\eq
\frac{d^{2} q_{\HO}}{dz^{2}} + \omega^{2} q_{\HO} = 0
\en
The time-dependent canonical transformation (\ref{eq:canonicaltransformation})
is a change of variables in the phase-space path integral over paths 
$\cQ(z)$ periodic in imaginary time.
So the imaginary time path integral is equivalent to that of a 
ordinary time-independent harmonic oscillator
giving an equilibrium thermal density matrix
which is thermodynamically stable.

\section{Numerical evidence for Property {\bf P}}

Property {\bf P} is the condition that the hermitian matrix $\cH(t)$
is positive definite on the physical phase space $\cV_{\phys}(t)$
for some $t\notin \{2mK\}$
and therefore for all such $t$.
\eq
\cH(t) = \Omega (\Moni(t)  -1) >0
\en
The thermal state for a given value of $\epsilon$ is stable
if and only if
property {\bf P} holds for all $(j_{L},j_{R}) \ne (0,0)$,
i.e.~for {\bf ode} $\mathbf{2}$,
$\;\mathbf{3}_{j}$, $\;\mathbf{2}_{j}$, 
and $\;\mathbf{1}_{j}$
for all values of the parameter $j$.

I do not know how to prove property {\bf P} 
for a given ode.
Instead I assemble strong numerical evidence
that property {\bf P} holds for {\bf ode} $\mathbf{2}$,
$\;\mathbf{3}_{j}$, $\;\mathbf{2}_{j}$, 
and $\;\mathbf{1}_{j}$
for all values of $j$
for all $\epsilon$ in the range $0\le\epsilon \le 
1/\sqrt2$  which is $1/2 \le k^{2} \le 1$.

The numerical calculations are done at $t=K$ because the function 
$\cn(K+\tau i,k)$ has symmetry properties that
allow for more robust numerical integration.
For each ode, the strategy is to check property {\bf P} numerically
on a finite sample of points in the parameter space.
For each point in the sample, property {\bf P} is found to hold.
Then $\cM_{i}(t)$ is diagonalized
and the minimum frequency $\omega_{\min}$ is calculated,
\eq
0< \omega_{\min} = \min \{\omega_{a}\}
\en
the numerical evidence shows
that $\omega_{\min}$ is bounded away from 0 throughout the parameter 
space
so  $\cH(t)$ is positive everywhere.
Property {\bf P} holds.

The numerical calculations are shown in 
\cite{Friedan2022:StabilitySuppMat}.

\vskip2ex
\noindent
{\bf ode }$\mathbf{2}$ 

The only parameter in ode $\mathbf{2}$ is $\epsilon$.
$N_{\phys}=1$  so there is only one frequency, $\omega_{\min}=\omega_{1}$.
Property {\bf P} is tested for
a discrete set of values of $\epsilon$.
At each  $\epsilon$
property {\bf P} holds  and
$\omega_{1}= \epsilon$ to the numerical precision of the calculation.
So property {\bf P} holds for all $\epsilon$.
The identity $\omega_{1} = \epsilon$
suggests that  ode $\mathbf{2}$ can be integrated analytically.

\vskip2ex
\noindent
{\bf odes$\;\mathbf{3}_{j}$, $\mathbf{2}_{j}$, $\mathbf{1}_{j}$}

Figures~\ref{fig:3j_finite}--\ref{fig:1j_asymp}
show the numerical data,
plotting $\omega_{\min}/p$ against $p$.
The calculations are done only 
for $j>3/2$ because the each ode is invariant under $j\rightarrow -j$,
$z\rightarrow z+2K'i$ since $\cn(z+2K'i,k) = - \cn(z,k)$.

The first graph for each ode
shows $(p,\omega_{\min}/p)$ at various $\epsilon$ for the lowest values of
$j$, $j=3/2,\,2,\,5/2,\,\ldots$.
The second graph shows the asymptotic limit $\epsilon\rightarrow 0$ 
with $p$ fixed.
The small $j$ data matches on to the asymptotic regime.
The two graphs combined show that $\omega_{\min}/p$ is
bounded away from 0 for all $\epsilon$,  certainly for
the physically interesting values $\epsilon<10^{-27} $.
So property {\bf P} holds.

Figure~\ref{fig:1j_smallp} clarifies the $\epsilon\rightarrow 0$  limit
at fixed $j$ for ode $\mathbf{1}_{j}$.
For each of the three ode series,
\eq
\omega_{\min} \xrightarrow[\epsilon\rightarrow 0]{} A \sigma
\en
for some constant $A$.  The parameters $\sigma$ and $p$ are related by
\eq
\frac{\sigma^{2}}{p^{2}} = \left\{
\begin{array}{c@{\qquad}l}
\frac1{6} & \text{\bf ode } \mathbf{3}_{j}\\[1ex]
\frac14 & \text{\bf ode } \mathbf{2}_{j}\\[1ex]
\frac{j^{2}}{4j^{2}+3} & \text{\bf ode } \mathbf{1}_{j}
\end{array}
\right.
\en
so
\eq
\frac{\omega_{\min}^{2}}{\omega_{\min,\asymp}^{2}} 
\quad \xrightarrow[\epsilon\rightarrow 0]{}\quad
\left\{
\begin{array}{c@{\qquad}l}
1 & \text{\bf ode } \mathbf{3}_{j}\\[1ex]
1 & \text{\bf ode } \mathbf{2}_{j}\\[1ex]
\frac{4j^{2}}{4j^{2}+3} & \text{\bf ode } \mathbf{1}_{j}
\end{array}
\right.
\en
For ode $\mathbf{1}_{j}$ the $\epsilon\rightarrow 0$ limit at fixed 
$j$ does not agree with the asympotic limit holding $p$ fixed.
This is an artifact of the parametrization by $p$.

\afterpage{%
\begin{figure}
\begin{center}
\captionsetup{justification=centering,margin=0.0\linewidth,skip=0pt}
\includegraphics[scale=1]{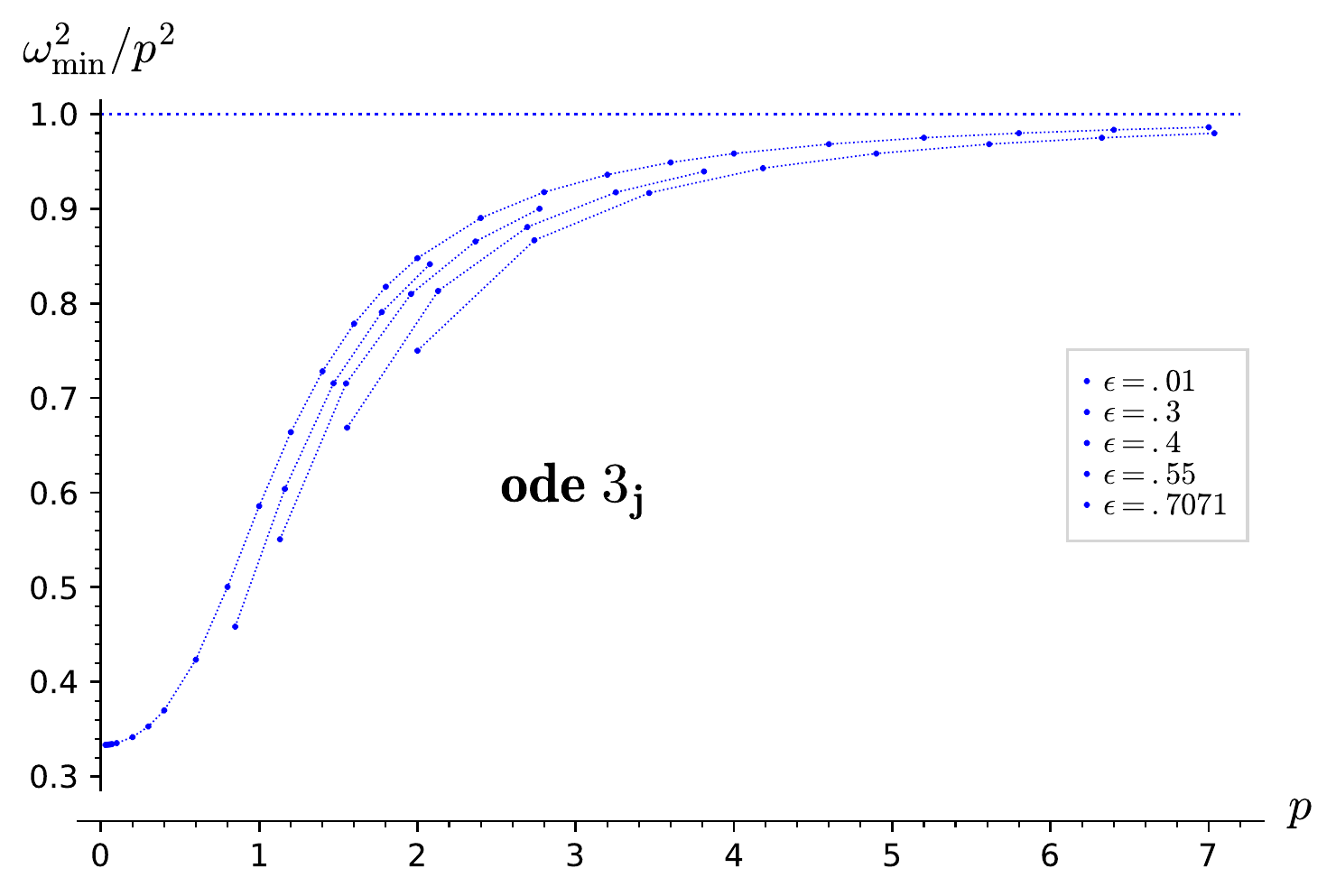}
\caption{
\label{fig:3j_finite}
The curves are ordered by increasing $\epsilon$ from left to right.
The dots are $j=3/2,\,2,\,5/2,\,\ldots$.  The connecting lines are 
interpolations.
}
\end{center}
\end{figure}

\begin{figure}
\begin{center}
\captionsetup{justification=centering,margin=0.0\linewidth,skip=0pt}
\includegraphics[scale=1]{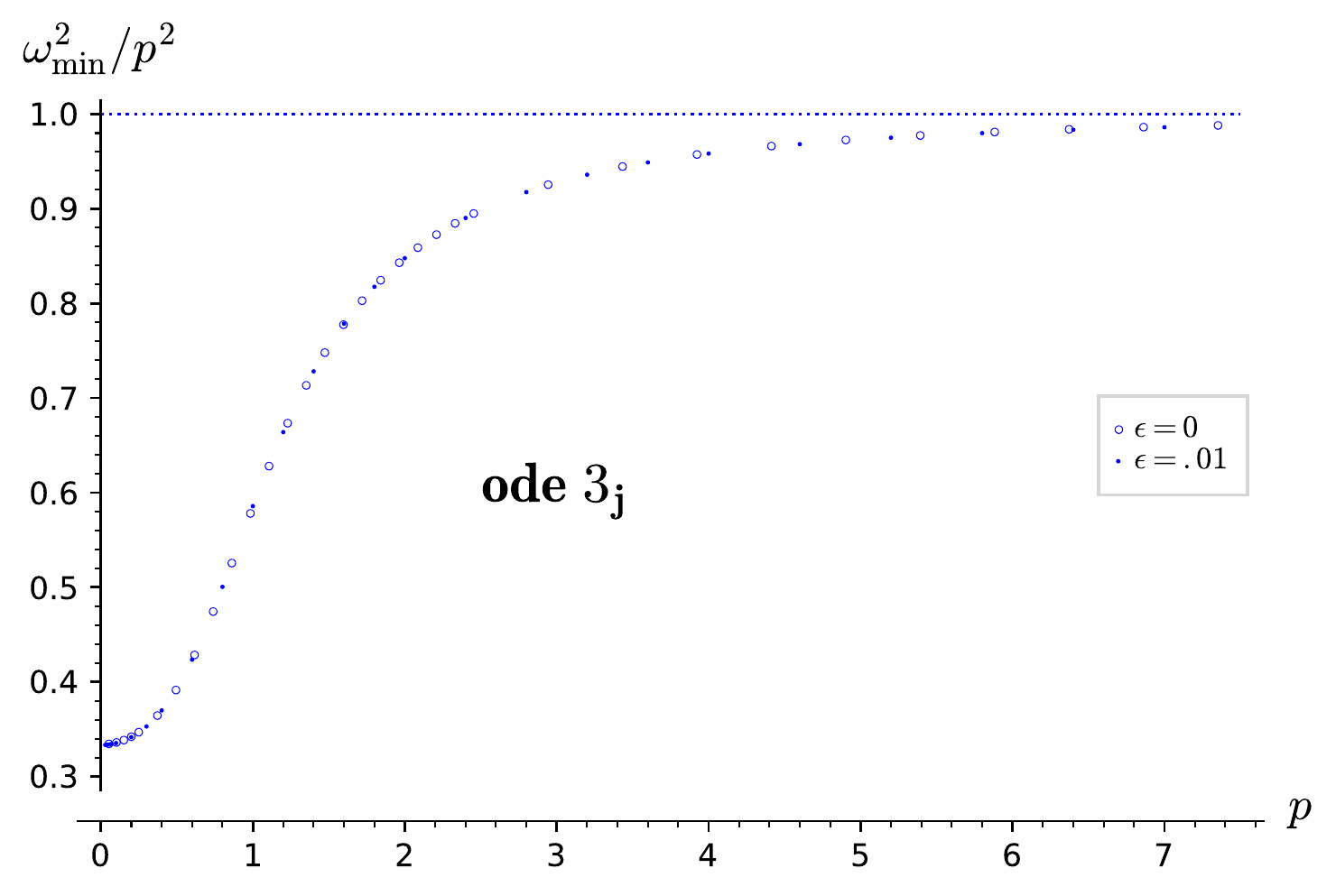}
\caption{
\label{fig:3j_asymp}
The open dots show the asymptotic limit $\epsilon\rightarrow 0$ with 
$p$ fixed.  The closed dots are the $\epsilon=.01$ data.
}
\end{center}
\end{figure}

\clearpage

\begin{figure}
\begin{center}
\captionsetup{justification=centering,margin=0\linewidth,skip=0pt}
\includegraphics[scale=1]{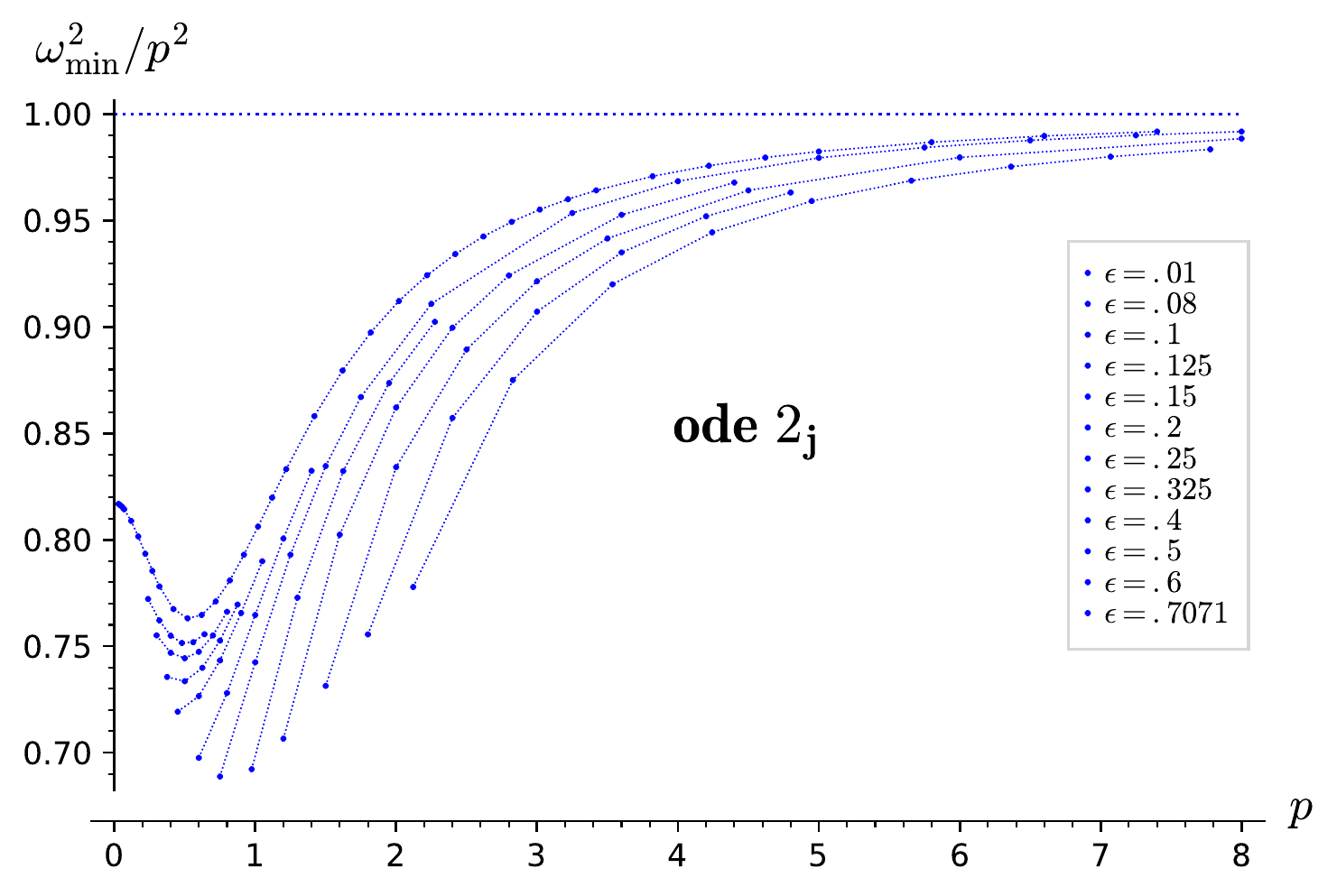}
\caption{
\label{fig:2j_finite}
The curves are ordered by increasing $\epsilon$ from left to righ.
The dots are $j=3/2,\,2,\,5/2,\,\ldots$.  The connecting lines are 
interpolations.
}
\end{center}
\end{figure}

\begin{figure}
\begin{center}
\captionsetup{justification=centering,margin=0\linewidth,skip=0pt}
\includegraphics[scale=1]{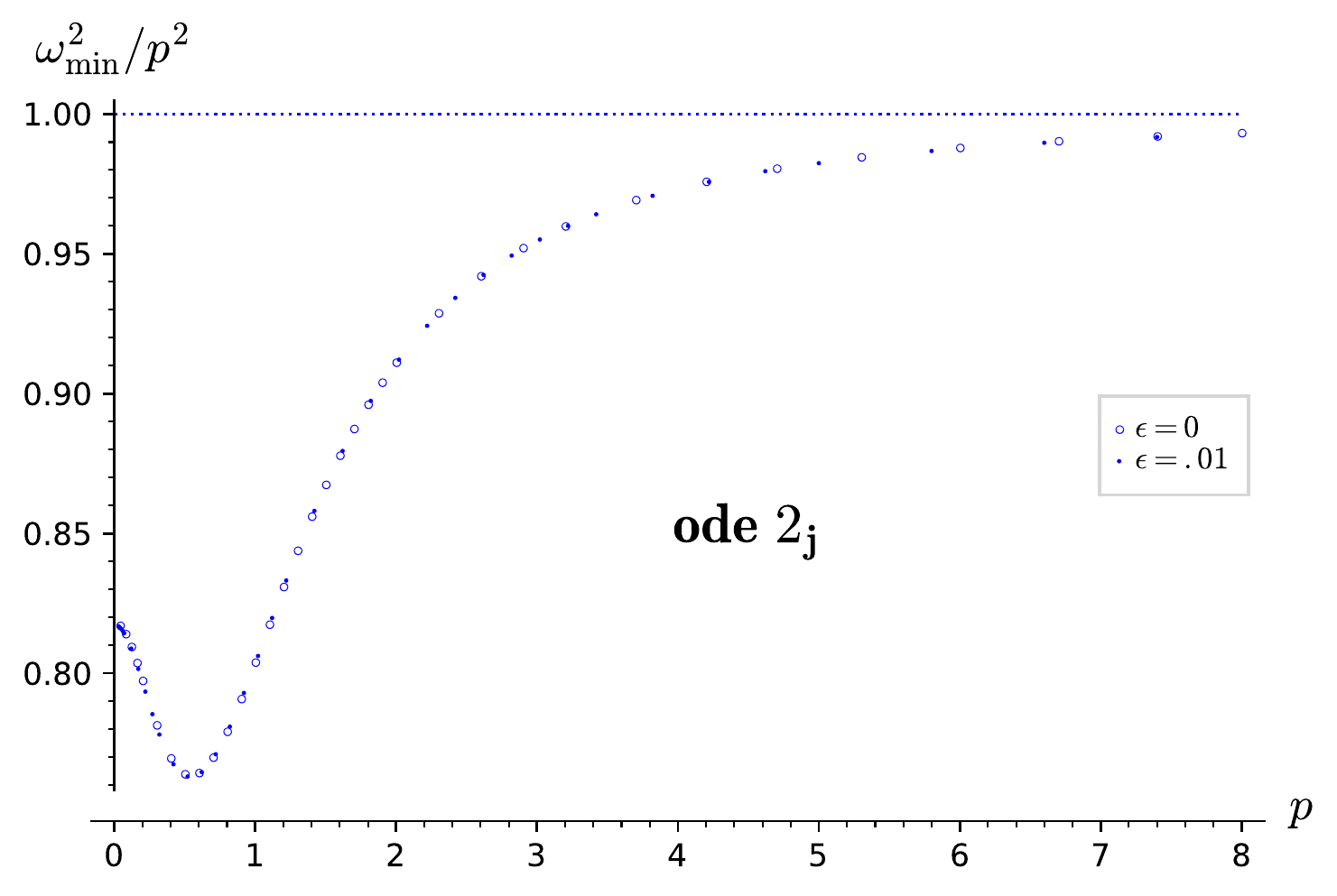}
\caption{
\label{fig:2j_asymp}
The open dots show the asymptotic limit $\epsilon\rightarrow 0$ with 
$p$ fixed.  The closed dots are the $\epsilon=.01$ data.
}
\end{center}
\end{figure}

\begin{figure}
\begin{center}
\captionsetup{justification=centering,margin=0\linewidth,skip=0pt}
\includegraphics[scale=1]{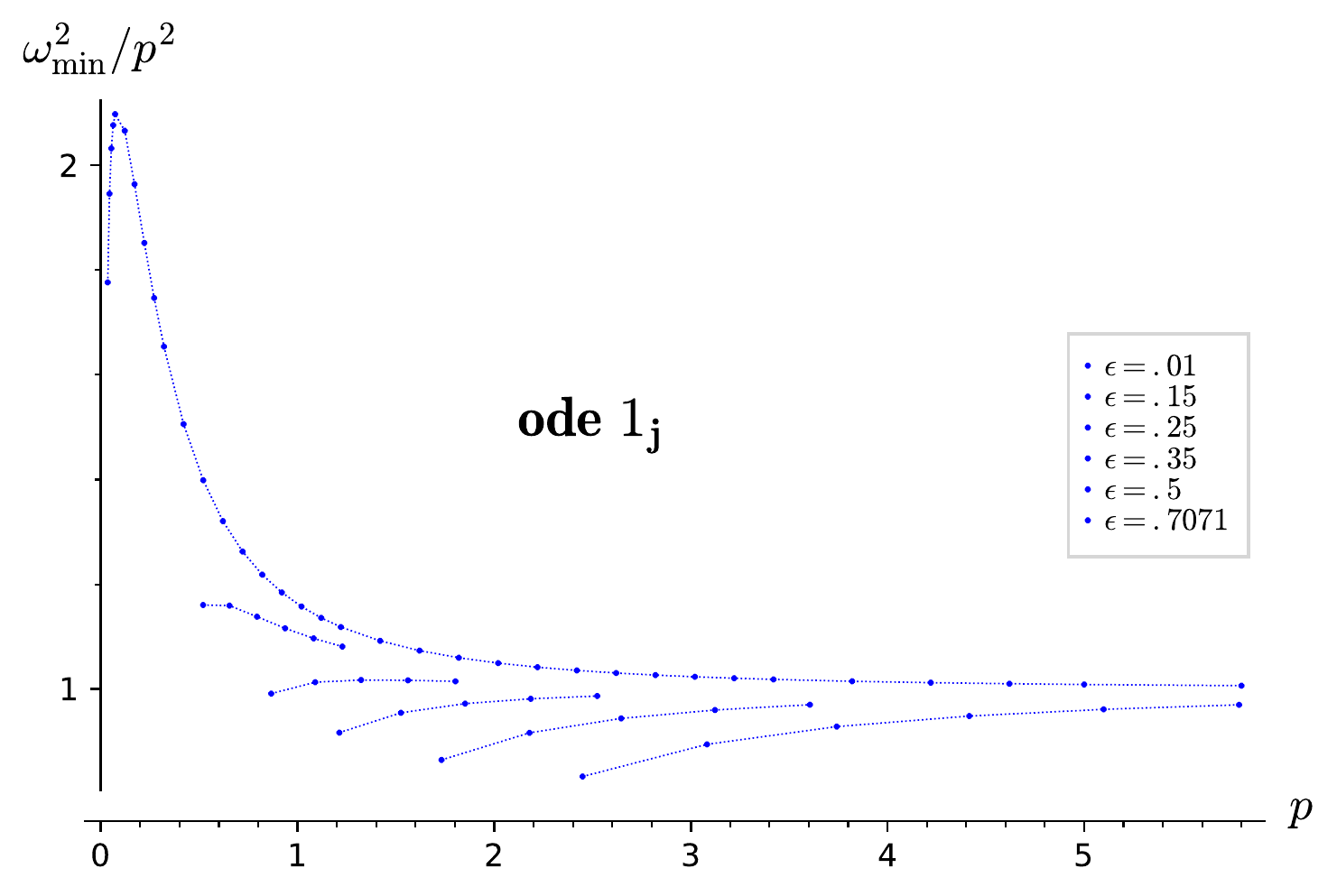}
\caption{
The curves are ordered by increasing $\epsilon$ from left to righ.
The dots are $j=3/2,\,2,\,5/2,\,\ldots$.  The connecting lines are 
interpolations.
}
\label{fig:1j_finite}
\end{center}
\end{figure}

\begin{figure}
\begin{center}
\captionsetup{justification=centering,margin=0\linewidth,skip=0pt}
\includegraphics[scale=1]{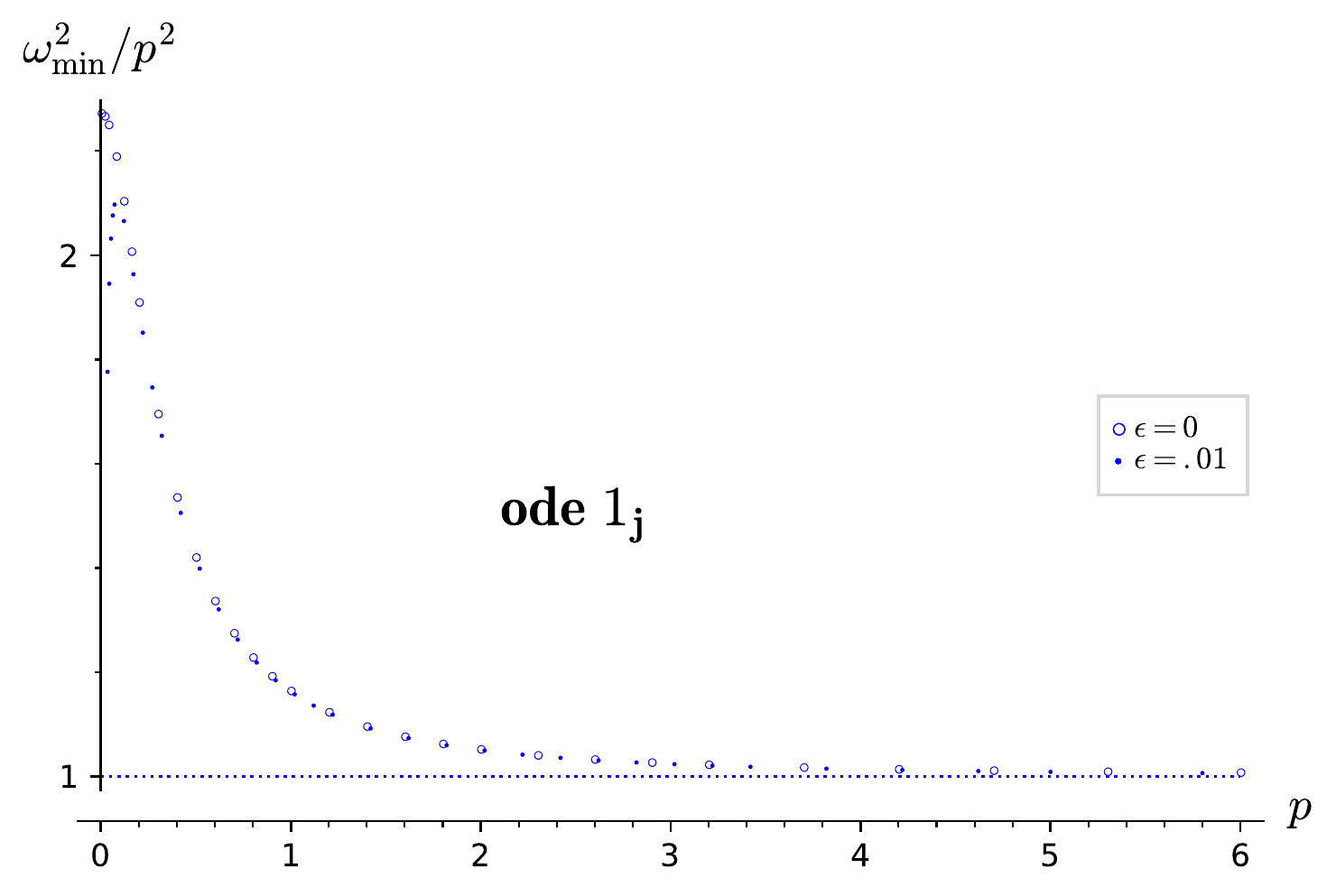}
\caption{
The open dots show the asymptotic limit $\epsilon\rightarrow 0$ with 
$p$ fixed.  The closed dots are the $\epsilon=.01$ data.
}
\label{fig:1j_asymp}
\end{center}
\end{figure}

\clearpage

\begin{figure}
\begin{center}
\captionsetup{justification=centering,margin=0\linewidth,skip=0pt}
\includegraphics[scale=1]{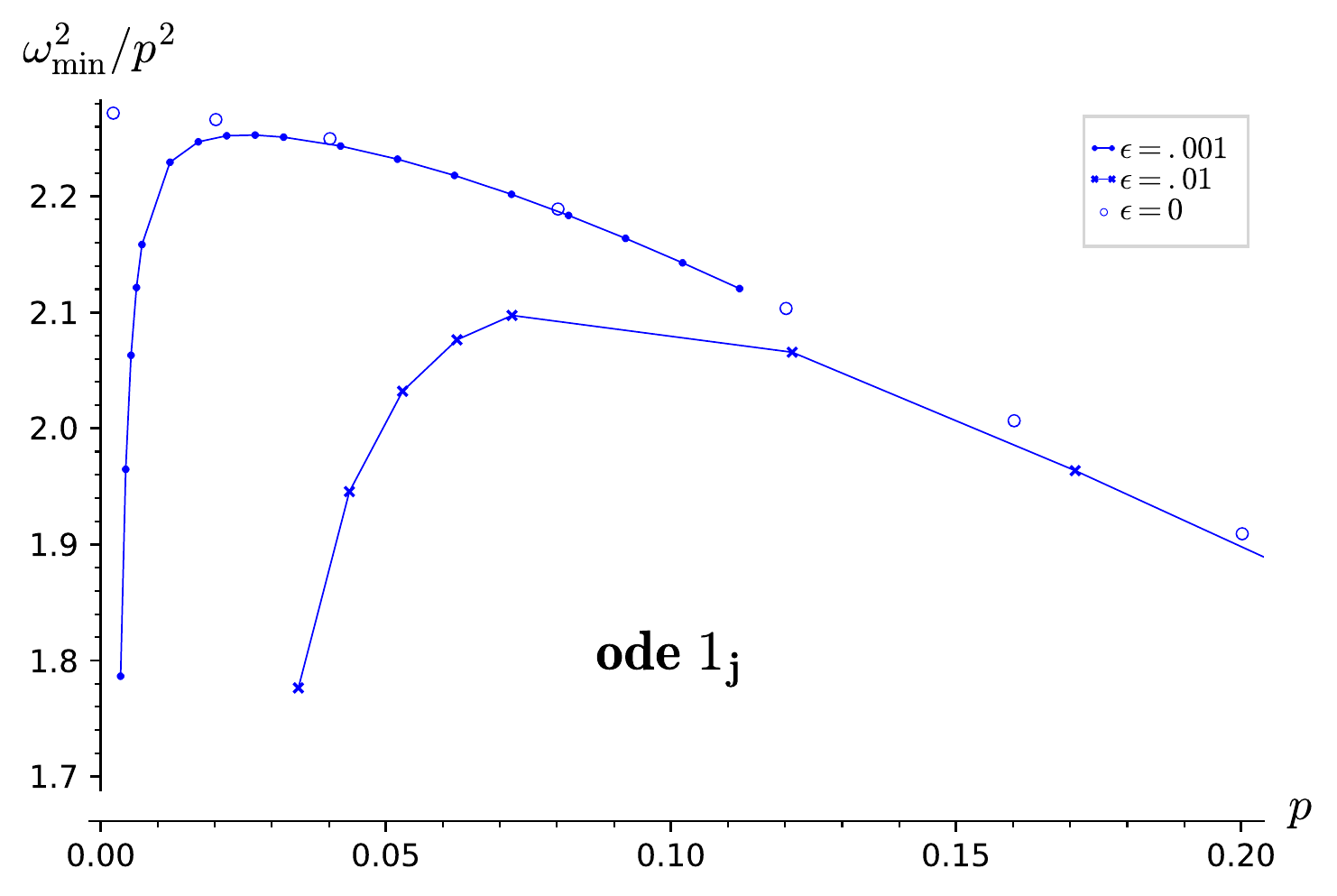}
\caption{
The approach to the asymptotic limit at small $p$.
}
\label{fig:1j_smallp}
\end{center}
\end{figure}
\clearpage

}

\section{Zero-mode integral}

\subsection{Observables}

The classical solution $b(z)$ defines a trajectory $z\mapsto 
\cB(z)$ in the  complexified phase space 
of the $\SU(2)$ gauge theory on $S^{3}$.
The periodicities of the classical solution imply that
$\cB(z)$ is a function on 
the complex plane modulo the lattice of periods,
the complex torus
\eq
\bbT = \Complexes/\bbL
\qquad
\bbL = \{ m(2K + 2K'i) + n(2K-2K'i) \colon m,n\in\Integers \}
\en
The poles of $\cB(z)$ are at $z\in \bbL \pm K'i$.
Away from the poles,
$\cB(z)$ takes values in
the complex phase space.
$\cB(z)$ can be defined at the poles
by adding points at infinity to the complex phase space.

The spaces $\cV_{\phys}(z)$ 
are the infinitesimal perturbations of $\cB(z)$ 
modulo the gauge symmetry.
Identifying the torus $\bbT$ with the classical trajectory,
the spaces $\cV_{\phys}(z)$ form a vector bundle over $\bbT$.
An observable --- a linear function of the gauge field fluctuations 
--- is an analytic section $\cF$ of the dual vector bundle,
\eq
\cF(z) \in \cV_{\phys}(z)^{*}
\en
which is nonsingular  away from the poles of $\cB(z)$.
The observable operator is
\eq
\cO(z)=
\cF(z)\, \cQ(z)
= \cF_{a}(z) \cQ^{a}(z) 
\en
$z$ is playing two roles here.  It is the complex time in $\cO(z)$ 
and $\cQ(z)$.  In $\cF(z)$, it is the complex coordinate on 
the classical trajectory in phase space.

The imaginary time gaussian 2-point expectation value of observables
\eq
\cO_{1}(z)=\cF_{1}(z)\, \cQ(z)
\qquad
\cO_{2}(z)=\cF_{2}(z)\, \cQ(z)
\en
is
\eqa
\expval{\cO_{2}(z_{2})\,\cO_{1}(z_{1})}_{t}
&= \cF_{2}(z_{2}) \expval{\cQ(z_{2}) \,\cQ(z_{1})^{t}} \cF_{1}(z_{1})^{t}
\\[1ex]
&= \cF_{2}(z_{2})\cP_{m}(z_{2},t) \cG(t,t) \cP_{m}(t,z_{1})^{t} \cF_{1}(z_{1})^{t}
\ena
\eq
t=t_{1}=t_{2}
\qquad
2mK < t < 2(m+1)K
\en
The vertical path $C$ through $t$ is in the simply connected 
region $R_{m}$ so
$\cP_{C}(z,t)=\cP_{m}(z,t)$.

The 2-point function has periodicity properties
\eqa
\expval{\cO_{2}(z_{2})\,\cO_{1}(z_{1})}_{t} &= \expval{\cO_{2}(z_{2}+4K'i)\,\cO_{1}(z_{1}+4K'i)}_{t}
\\[1ex]
&= \expval{\cO_{2}(z_{2}+2K+2K'i)\,\cO_{1}(z_{1}+2K+2K'i)}_{t+2K}
\\[2ex]
\label{eq:gaussianKMS}
\expval{\cO_{2}(t+4K'i)\,\cO_{1}(z_{1})}_{t} 
&= \expval{\cO_{1}(z_{1})\,\cO_{2}(t)}_{t} 
\ena

\subsection{Integrate over time translations}

Expectation values in the thermal state
are the integrals of the gaussian expectation values
over the time-translation zero-mode.
The thermal 2-point expectation value is
\eq
\expval{\cO_{2}(z_{2})\,\cO_{1}(z_{1})}
= \frac1{4K'}\int_{0}^{4K'} d\tau
\;
\expval{\cO_{2}(z_{2}+\tau i)\,\cO_{1}(z_{1}+\tau i)}_{t}
\en
Change variable to $z_{0}= z_{1}+\tau i$.
Recall that $C'_{t}$ is the vertical path from $0$ to $4K'i$.
\eq
\label{eq:thermaltwopoint}
\expval{\cO_{2}(z_{2})\,\cO_{1}(z_{1})}
= \frac1{4K' i}\int_{C'_{t}} dz_{0}
\;
\expval{\cO_{2}(z_{21}+z_{0})\,\cO_{1}(z_{0})}_{t}
\qquad
z_{21} = z_{2}-z_{1}
\en
Recall that $z_{1}$ and $z_{2}$ are both on $C'_{t}$ with
$0\le \tau_{1} \le \tau_{2}\le 4K'$,
so $z_{21} = \tau_{21}i$ with $0\le \tau_{21}\le 4K'$.
Later we will analytically continue the thermal expectation value to 
the complex $z_{21}$ plane in order to construct the real time 
2-point expectation values.

\subsection{The thermal state is independent of $t$}
\label{sec:thermalstateindoft}

Given that $z_{21} = \tau_{21}i$,
the integrand in (\ref{eq:thermaltwopoint}) is
a nonsingular analytic function of $z_{0}$ 
in the vertical strip  $2mK < t_{0} < 2(m+1)K$.
For any $C'_{t'}$ in the same vertical strip,
\eq
\int_{C'_{t'}-C'_{t}} dz_{0} 
\;
\expval{\cO_{2}(z_{21}+z_{0})\,\cO_{1}(z_{0})}_{t}
=
\int_{t}^{t'} - \int_{t+4K'i}^{t'+4k'i}
\;
\expval{\cO_{2}(z_{21}+z_{0})\,\cO_{1}(z_{0})}_{t}
=0 
\en
because the integrand is periodic under $z_{0}\rightarrow z_{0}+4K'i$.
So the rhs of (\ref{eq:thermaltwopoint}) is constant in $t$ within 
each vertical strip.
The periodicity $z\rightarrow z+2K+2K'i$ gives
\eqa
\frac1{4K' i}&\int_{C'_{t}} dz_{0}
\;
\expval{\cO_{2}(z_{21}+z_{0})\,\cO_{1}(z_{0})}_{t}
\\[1ex]
&= \frac1{4K' i}\int_{C'_{t}} dz_{0}
\;
\expval{\cO_{2}(z_{21}+z_{0}+2K+2K'i)\,\cO_{1}(z_{0}+2K+2K'i)}_{t}
\\[1ex]
&= 
\frac1{4K' i}\int_{C'_{t+2K}} dz_{0}'
\;
\expval{\cO_{2}(z_{21}+z_{0}')\,\cO_{1}(z_{0}')}_{t}
\ena
so the rhs of (\ref{eq:thermaltwopoint}) is the same on neighboring 
strips.  So it is independent of $t$.
The thermal state is well defined, independent of the choice of $t$.

\subsection{KMS condition}

The last periodicity identity in (\ref{eq:gaussianKMS}) gives
\eqa
\expval{\cO_{2}(z_{2})\,\cO_{1}(z_{1})}
& = \frac1{4K'}\int_{0}^{4K'} d\tau
\;
\expval{\cO_{2}(z_{2}+\tau i)\,\cO_{1}(z_{1}+\tau i)}_{t}
\\[1ex]
&=
\frac1{4K'}\int_{0}^{4K'} d\tau
\;
\expval{\cO_{1}(z_{1}+\tau i+4K'i)\,\cO_{2}(z_{2}+\tau i)}_{t}
\ena
which is the KMS condition
\eq
\expval{\cO_{2}(z_{2})\,\cO_{1}(z_{1})}
= \expval{\cO_{1}(z_{1}+4K'i)\,\cO_{2}(z_{2})}
\en

\section{Analytic continuation to real time}

The euclidean signature formulation of thermodynamic stability
should imply real time stability.
As a  step towards checking this,
the analytic continuation
to real time of
the imaginary time two-point function
is constructed, leaving for later the
study of its large time asymptotic behavior.

Fix $m=0$ and simply-connected region $R_{0}$.
For every $t$ in the interval $0< t < 2K$
the integrand in (\ref{eq:thermaltwopoint}) is analytic and 
nonsingular in both $z_{0}$ 
and in $z_{21}$ as long as
$0< t+t_{21} < 2K$.
Thus for each such $t$ there is an analytic continuation of the 
two-point expectation value to the strip
$-t < t_{21} < 2K-t$.
The periodic paths $C_{t}'$ for $0< t < 2K$ are all deformable to 
each other,
so all of these analytic continuations agree on the overlaps.
So $\expval{\cO_{2}(z_{21})\,\cO_{1}(0)}$ is analytic in the strip
$-2K < t_{21} < 2K$.

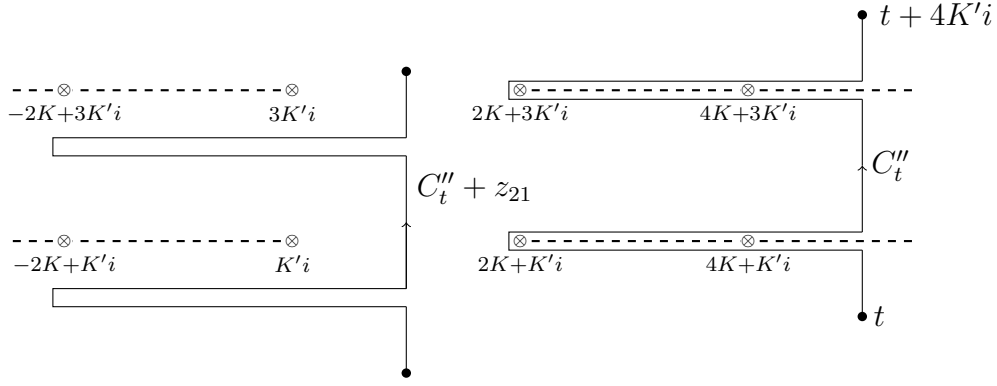
\begin{figure}
\begin{center}
\begin{tikzpicture}[baseline=2ex,x=1.5cm,y=1cm]
\foreach \y in {-1,1}
{
\node at (-2,\y) {$\scriptscriptstyle \otimes$};
\node at (0,\y) {$\scriptscriptstyle \otimes$};
\node at (2,\y) {$\scriptscriptstyle \otimes$};
\node at (4,\y) {$\scriptscriptstyle \otimes$};
\draw[dashed,thick] (4.1,\y) -- (5.5,\y);
\draw[dashed,thick] (2.1,\y) -- (3.93,\y);
\draw[dashed,thick] (-0.1,\y) -- (-1.93,\y);
\draw[dashed,thick] (-2.1,\y) -- (-2.5,\y);
}
\node at (0,-1-.3){$\scriptstyle K'i$};
\node at (0,1-.3){$\scriptstyle 3 K'i$};
\node at (2,-1-.3){$\scriptstyle 2K+K'i$};
\node at (2,1-.3){$\scriptstyle 2K+3K'i$};
\node at (4,-1-.3){$\scriptstyle 4K+K'i$};
\node at (4,1-.3){$\scriptstyle 4K+3K'i$};
\node at (-2,-1-.3){$\scriptstyle -2K+K'i$};
\node at (-2,1-.3){$\scriptstyle -2K+3K'i$};
\draw[] (5,-2) -- (5,-1.12) ;
\draw[] (5,-1.12) -- (1.9,-1.12) -- (1.9,-1+.12) -- (5,-1+.12);
\draw[->] (5,-1+.12) -- (5,0);
\draw[] (5,0) -- (5,1-.12);
\draw[] (5,2-1.12) -- (1.9,2-1.12) -- (1.9,2-1+.12) -- (5,2-1+.12);
\draw[] (5,1.12) -- (5,2);
\draw[] (-4 +5,-0.75-2) -- (-4 +5,-0.75-1.12) ;
\draw[] (-4 + 5,-0.75-1.12) -- (-4 + 1.9,-0.75-1.12) -- (-4 + 1.9,-0.75-1+.12) -- (-4 + 5,-0.75-1+.12);
\draw[->] (-4 + 5,-0.75-1+.12) -- (-4 + 5,-0.75+0);
\draw[] (-4 + 5,-0.75-1+.12) -- (-4 + 5,-0.75+1-.12);
\draw[] (-4 + 5,-0.75+2-1.12) -- (-4 + 1.9,-0.75+2-1.12) -- (-4 + 1.9,-0.75+2-1+.12) -- (-4 + 5,-0.75+2-1+.12);
\draw[] (-4+5,-0.75+1.12) -- (-4+5,-0.75+2);
\node at (5+0.15,-2){$\textstyle  t$};
\node at (5+0.65,2){$t+4K'i$};
\node at (-4 +5,-0.75+-2) [circle,fill,inner sep=1.25pt]{};
\node at (-4 +5,-0.75+2) [circle,fill,inner sep=1.25pt]{};
\node at (5,-2) [circle,fill,inner sep=1.25pt]{};
\node at (5,2) [circle,fill,inner sep=1.25pt]{};
\node at (1.6,-0.3+0){${\textstyle  C''_{t}}+z_{21}$};
\node at (4+1.25,0){$\textstyle  C''_{t}$};
\end{tikzpicture}
\end{center}
\caption{
For $t\ge 2K$ the periodic path $C''_{t}$ goes from
$t$ to $t+4K'i$ staying within $R_{0}$.
The path $C''_{21} + z_{21}$
also stays within $R_{0}$ 
if $-2K < t+t_{21} < 2K$
unless $\tau_{21} \in 2 K' \Integers$.
}
\label{fig:Cdoubleprime}
\end{figure}
To analytically continue beyond this strip,
let $C''_{t}$ for $t\ge 2K$
be the periodic path within $R_{0}$ from $t$ to $t+4K'i$
shown in Figure \ref{fig:Cdoubleprime}.
For $t\le -2k$,
let $C''_{t}$ be the reflection of $C''_{-t}$ through 
the imaginary axis, $C''_{t} = - \overline{C''_{-t}}$.
All the periodic paths $C'_{t}$, $-2K<t<2K$, and $C''_{t}$, $2K \le |t|$,
are deformable to each other.
Then, for $2K \le |t|$,
\eq
\expval{\cO_{2}(z_{2})\,\cO_{1}(z_{1})}
= \frac1{4K' i}\int_{C''_{t}} dz_{0}
\;
\expval{\cO_{2}(z_{21}+z_{0})\,\cO_{1}(z_{0})}_{t}
\en
is analytic in $z_{21}$ for
\eq
-2K -t < t_{21} < 2K-t
\quad\text{and}\quad
\tau_{21} \not\in 2 K' \Integers
\en
because these are the conditions that
$z_{0}+z_{21}\in R_{0}$ for all $z_{0}$ in $C''_{t}$.
Again, all the locally analytic constructions agree on the overlaps 
by contour deformation.

Therefore $\expval{\cO_{2}(z_{2})\,\cO_{1}(z_{1})}$
\begin{enumerate}
\item is time-translation invariant,
\eq
\expval{\cO_{2}(z_{2})\,\cO_{1}(z_{1})} = 
\expval{\cO_{2}(z_{21})\,\cO_{1}(0)}
\qquad
z_{21}=z_{2}-z_{1}=t_{21}+\tau_{21}i
\en
\item is analytic in $z_{21}$ in the region
\eq
-2K < t_{21} < 2K
\quad
\text{or}
\quad
|t_{21}| \ge 2K,\;\;
\tau_{21} \not\in 2K' \Integers
\en
shown in Figure~\ref{fig:z12analytic}.
\item satisfies the KMS periodicity condition
\eqa
\expval{\cO_{2}(z_{2})\,\cO_{1}(z_{1})}
&= \expval{\cO_{1}(z_{1}+4K'i)\,\cO_{2}(z_{2})}
\\[1ex]
\expval{\cO_{2}(z_{21})\,\cO_{1}(0)}
&=\expval{\cO_{1}(4K'i-z_{21})\,\cO_{2}(0)}
\ena
\end{enumerate}
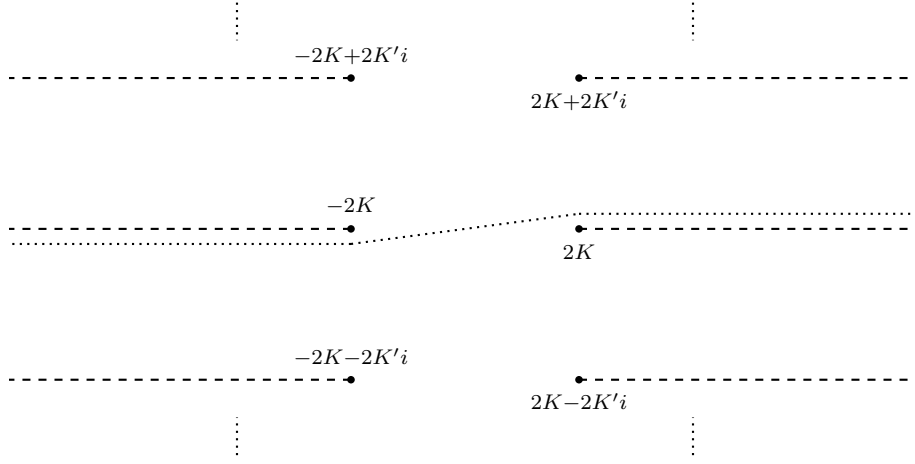
\begin{figure}
\begin{center}
\begin{tikzpicture}[baseline=2ex,x=1.5cm,y=1cm]
\foreach \y in {-2,0,2}
{
\draw[dashed,thick] (1,\y) -- (4,\y);
\draw[dashed,thick] (-1,\y) -- (-4,\y);
\node at (1,\y) [circle,fill,inner sep=1pt]{};
\node at (-1,\y) [circle,fill,inner sep=1pt]{};
}
\node at (1,0-.3){$\scriptstyle 2K$};
\node at (1,2-.3){$\scriptstyle 2K+2K'i$};
\node at (1,-2-.3){$\scriptstyle 2K-2K'i$};
\node at (-1,0+.3){$\scriptstyle -2K$};
\node at (-1,2+.3){$\scriptstyle -2K+2K'i$};
\node at (-1,-2+.3){$\scriptstyle -2K-2K'i$};
\draw[dotted,thick] (-1,-0.2) -- (-4,-0.2);
\draw[dotted,thick] (1,0.2) -- (4,0.2);
\draw[dotted,thick] (-1,-0.2) -- (1,0.2);
\draw[dotted,thick] (2,3) -- (2,2.5);
\draw[dotted,thick] (2,-3) -- (2,-2.5);
\draw[dotted,thick] (-2,3) -- (-2,2.5);
\draw[dotted,thick] (-2,-3) -- (-2,-2.5);
\end{tikzpicture}
\end{center}
\caption{
The domain of analyticity of $\expval{\cO_{2}(z_{21})\,\cO_{1}(0)}$.
Real time is the dotted line.
}
\label{fig:z12analytic}
\end{figure}
The analytic continuation to real time is
from above the real axis for $t_{21}>2K$ and from below the axis for 
$t_{21}<-K$, as shown in Figure \ref{fig:z12analytic}.

\section{Comments and questions}

\subsection{A geometric proof?}

There ought to be a geometric proof of thermodynamic 
stability directly from the self-adjointness of the Yang-Mills hamiltonian.
The natural mathematical setting is the 
complexification of the phase-space of the $\SU(2)$ Yang-Mills theory on $S^{3}$.
The $\Spin(4)$-invariant $\SU(2)$ gauge solution 
for each value of the Yang-Mills energy is an elliptic curve, a torus, in the 
complex phase-space.
The Yang-Mills energy parametrizes the moduli space of elliptic curves.
The collection of all the solutions comprises
a map from the universal elliptic curve to the complex phase-space.
A rigorous construction of this map
requires adding points at infinity to the 
complex phase-space to accommodate the poles in the classical 
solution.  The complex gauge field $B_{i}^{a}(\hat x)$ at each point $\hat x$ 
lives in a complex 9-dimensional vector space.  Perhaps it is enough 
to projectify each of these vector spaces so the gauge field at each 
$\hat x$ lives in $\Complexes \mathbf{P}^{9}$.
Once the projectification of the complex phase-space is constructed,
the Yang-Mills theory has to be extended to the projectification.
If this can be done, it should be possible to pull back the hamiltonian 
structure to the universal curve and construct the thermal state.

\subsection{Other natural states?}

There is a certain arbitrariness in defining the thermal state by 
the imaginary time periodicity $z \rightarrow z+4K'i$.
Why not the state defined by the periodicity $z\rightarrow z+2K+2K'i$?
Are these different states?  Is there more than one natural
thermodynamically stable state?

\phantomsection
\section*{Acknowledgments}
\addcontentsline{toc}{section}{\numberline{}Acknowledgments}
This work was supported by the Rutgers New High Energy Theory Center
and by the generosity of B. Weeks.
I am grateful
to the  Mathematics Division of the 
Science Institute of the University of Iceland
for its hospitality.

\bibliographystyle{ytphys}
\raggedright
\phantomsection
\bibliography{Literature}
\addcontentsline{toc}{section}{\numberline{}References}

\end{document}